# Discovery of Orbital Ordering in $Bi_2Sr_2CaCu_2O_{8+x}$

Shuqiu Wang[1,2,3§], Niall Kennedy[1,4§], K. Fujita[5], S. Uchida[6], H. Eisaki[7],

P.D. Johnson[1,5], J.C. Séamus Davis[1,2,4,8] and S.M. O'Mahony[4]

1. Clarendon Laboratory, University of Oxford, Oxford, OX1 3PU, UK.
2. Department of Physics, Cornell University, Ithaca NY 14850, USA
3. H. H. Wills Physics Laboratory, University of Bristol, Bristol, BS8 1TL, UK
4. School of Physics, University College Cork, Cork T12 R5C, Ireland
5. CMPMS Department, Brookhaven National Laboratory, Upton, NY, USA
6. Department of Physics, University of Tokyo, Bunkyo, Tokyo 113-0011, Japan
7. Inst. of Advanced Industrial Science and Tech., Tsukuba, Ibaraki 305-8568, Japan.
8. Max-Planck Institute for Chemical Physics of Solids, D-01187 Dresden, Germany
§ These authors contributed equally to this project.

**The primordial ingredient of cuprate superconductivity is the $CuO_2$ unit cell. Here, theoretical attention usually concentrates on the intra-atom Coulombic interactions dominating the $3d^9$ and $3d^{10}$ configurations of each copper ion. However, if Coulombic interactions also occur between electrons of the $2p^6$ orbitals of each planar oxygen atom, spontaneous orbital ordering may split their energy levels. This long predicted intra-unit cell symmetry breaking should then generate an orbital ordered phase, for which the charge-transfer energy $\mathcal{E}$ separating the $2p^6$ and $3d^{10}$ orbitals is distinct for the two oxygen atoms. Here we introduce sublattice resolved $\mathcal{E}(r)$ imaging techniques to $CuO_2$ studies and discover intra-unit-cell rotational symmetry breaking of $\mathcal{E}(r)$, with energy-level splitting between the two oxygen atoms on the 50 meV scale. Spatially, this state is arranged in disordered Ising domains of orthogonally oriented orbital order that appear bounded by dopant ions, and within whose domain walls low energy electronic quadrupolar two-level systems occur. Overall, these data reveal a $Q$=0 orbitally ordered state that splits the energy levels of the oxygen orbitals by ~50 meV, in underdoped $CuO_2$.**

**Orbital Ordering in High Temperature Superconductors**

**1**  Unforeseen and unexplained among the quantum matter states of hole-doped $CuO_2$ is an electronic nematic phase[1]. Analyses of the three-orbital model[2-9] for the $CuO_2$ charge-transfer insulator have long predicted a candidate mechanism for the nematic state involving an energy splitting between the two planar-oxygen $2p^6$ orbitals within each unit cell. Although never observed, such effects should be identifiable experimentally as a difference in charge-transfer energy $\mathcal{E}(\boldsymbol{r})$ separating each oxygen $2p^6$ orbital from the relevant copper $3d^{10}$ orbital configuration. Moreover, if extant, such ordering of the oxygen $2p^6$ orbitals may be juxtaposed with iron-based high temperature superconductors where intra-unit-cell rotational symmetry breaking between the iron $d_{zx}$ and $d_{zy}$ orbitals[10-12] is pivotal[13,14].

**2**  Soon after the discovery of Fe-based high temperature superconductivity the profound significance of ordering in the iron $d_{zx}$ and $d_{zy}$ orbitals was identified[10-12], along with the subsequent realization that such orbital ordering (lifting energy degeneracy of $d_{zx}$ and $d_{zy}$ orbitals) was a fundamentally important phenomenon[13,14]. In these materials, such orbital ordering is pivotal for the tetragonal to orthorhombic phase transition into an electronic nematic phase[1,13]. This lifting of $d_{zx} : d_{zy}$ energy degeneracy certainly has profound global effects[13,14]: sequentially with falling temperature, the orbital ordering renders the electronic structure strongly nematic along with a strongly anisotropic antiferromagnetic state; its spectrum of magnetic fluctuations exhibits equivalently reduced symmetry[13]; finally a strongly anisotropic[14] and even orbital selective [15] form of high temperature superconductivity emerges. Hence, intra-unit-cell orbital ordering produces powerful, wide-ranging effects on the electronic structure, quantum magnetism, high temperature superconductivity and on the global phase diagram of the Fe-based high temperature superconductive materials[13,14, 16, 17]. However, although long anticipated[2-9], analogous intra-unit-cell orbital ordering for Cu-based high temperature superconductive materials, has never been observed.

**Charge Transfer Superexchange**

**3** In these materials, the planar Cu$^{++}$ ions are in the 3$d^9$ configuration with a singly occupied $d_{x^2-y^2}$ orbital, while the planar O$^{--}$ ions have filled 2$p^6$ orbitals. The Cu 3$d^{10}$ configuration, which is energetically disfavoured by the Coulomb energy $U$ required to double occupy each $d_{x^2-y^2}$ orbital, is separated from the intervening oxygen 2$p^6$ energy levels by the charge-transfer energy $\mathcal{E}$, as shown in Fig. 1d. While at half-filling the $d$-electrons are fully localized in a charge-transfer insulator state, introduction of holes into the oxygen 2$p^6$ orbitals radically alters the situation, necessitating a three-orbital Hamiltonian[18, 19] description

$$H = \sum_{i\alpha j\beta,\sigma} t_{ij}^{\alpha\beta} c_{i\sigma}^{\dagger\alpha} c_{j\sigma}^{\beta} + \sum_{i\sigma,\alpha} \varepsilon_\alpha n_{i\sigma}^\alpha + U \sum_i n_{i\uparrow}^d n_{i\downarrow}^d. \quad (1)$$

Here $\alpha, \beta$ label any of the three orbitals, $t_{ij}^{\alpha\beta}/\hbar$ are transition rates for electrons between orbitals $\alpha, \beta$ at sites $i, j$, $\varepsilon_\alpha$ are the orbital energies, and $n_{i\uparrow}^d, n_{i\downarrow}^d$ are the $d$-orbital occupancies. High temperature superconductivity is believed to emerge within this model driven by the charge-transfer superexchange pairing interactions between the Cu $d$-electrons (Methods Section A). In theory[1], however, many other ordered phases could emerge upon hole doping CuO$_2$, with a planar oxygen orbital-ordered phase[2-9] due to additional inter-oxygen repulsion terms $V_{pp} \sum_{ij\sigma\sigma'} n_{i\sigma}^p n_{j\sigma'}^p$, in Eqn. 1, being a prime example.

**4** One enigmatic phase that does emerge is the *pseudogap* (PG) regime. Its essential phenomenology[20,21] is that, for $T < T^*(p)$ and $p < p^* \approx 0.19$ (yellow shading Fig. 1a), the Fermi surface becomes partially gapped thus diminishing the spectrum of electronic states, the magnetic susceptibility, the electronic specific heat, the $c$-axis electrical conductivity, and the spatially-averaged density of electronic states. Strikingly, however, in this same region of the phase diagram there is pervasive evidence for a nematic state in which the electronic structure breaks 90°-rotational (C$_4$) symmetry at wavevector ***Q***=0, or equivalently within the CuO$_2$ unit cell. Experimental signatures of this nematic state have been very widely reported based on multiple techniques as summarized in Fig. 1a, and this nematic phenomenology is observed for $p < p^*$ in the Bi$_2$Sr$_2$CaCu$_2$O$_{8+x}$, Bi$_2$Sr$_2$CuO$_{6+x}$, Bi$_{2-z}$Pb$_z$Sr$_{2-y}$La$_y$CuO$_{6+x}$, YBa$_2$Cu$_3$O$_{7-x}$, La$_{2-x}$Ba$_x$CuO$_4$, and HgBa$_2$CuO$_{4+\delta}$ material families, thus strongly indicative of universality (Methods Section B). But no microscopic mechanism has yet been experimentally established for this CuO$_2$ nematic phase.

**Motivation for Cuprate Orbital Ordering**

*5*     Classically, a nematic metallic state may occur due to the Pomeranchuk instability of the Fermi surface [22]. But, for cuprates, theoretical analyses using a three-orbital band structure [2-9] predict the emergence of low energy electronic nematicity driven primarily by high energy orbital ordering. Specifically, when the Coulomb repulsions between electrons on nearest-neighbour O sites within the same unit cell are included in the strong-coupling limit [2] and in self-consistent mean-field theory [3], an orbitally ordered nematic phase is predicted. On the same basis, such CuO$_2$ orbital order is also predicted using diagrammatic perturbation theory [4], perturbative expansion [5], functional-renormalization-group techniques [7], Hartree-Fock based models [8], and in slave-boson mean-field theory [9]. Figure 1b represents conceptually such planar-oxygen orbital-order while Fig. 1c represents its schematic impact on the intra-unit-cell characteristics of charge-transfer energy $\mathcal{E}$. Here the $2p^6$ orbital of the oxygen atom along the CuO$_2$ $x$-axis (O$_x$ site) is separated from the upper Cu band by the charge transfer energy $\mathcal{E}_x$, while the notionally equivalent oxygen orbital along the $y$-axis (O$_y$ site) exhibits a different charge transfer energy $\mathcal{E}_y$. Our objective is then a direct search for such rotational symmetry breaking at the charge-transfer energy scale, through visualization of $\mathcal{E}$ within each CuO$_2$ unit cell.

**Visualization of Charge Transfer Energy Variations**

*6*     In practice, Bi$_2$Sr$_2$CaCu$_2$O$_{8+x}$ samples with hole-density $p \approx 0.17$ and $T_c = 88$ K are inserted into a spectroscopic imaging scanning tunneling microscope (SISTM) and cleaved in cryogenic ultra-high vacuum at $T$ = 4.2 K. The technique of imaging charge transfer energies in cuprates has been established by several STM studies [23-25] of variations in the $dI/dV$ spectra measured over the energy range -1.5 eV < $E$ < 2 eV. However, intra-unit-cell resolution imaging of the charge transfer energies have proven challenging due to the high tunneling junction resistance (and thus large tip-sample separation) required for such high voltage imaging. Here we overcome such challenges and introduce sublattice resolved [26-29] imaging of $\mathcal{E}(\boldsymbol{r})$ to $CuO_2$ studies. A typical topographic image, $T(\boldsymbol{r})$, of the terminal BiO surface from these studies, is shown in Fig. 2a. To visualize the intra-unit-cell structure of the charge-transfer energy scale, we use a recently developed technique[25] which analyzes high-voltage tunneling conductance spectra $g(\boldsymbol{r}, E) \equiv dI/dV(\boldsymbol{r}, eV)$ so as yield the spatial variations in $\mathcal{E}(\boldsymbol{r})$. Figure 2b shows an exemplary high-voltage $g(\boldsymbol{r}, V)$ spectrum, measured using junction-resistance $R_N \approx 85$ G$\Omega$ at a specific site $\boldsymbol{r}$ within the $CuO_2$ unit cell (yellow dot in Fig. 2a). By comparison to the spatially averaged $\overline{g(V)}$ in the same field-of-view (FOV) we see that they have different energy separation between the lower band (filled) and the upper band (empty) (compare red and blue arrows in Fig. 2b). Hence, by visualizing $g(\boldsymbol{r}, V)$ in the $V_{min}(-1.6 \text{ V}) \leq V \leq V_{max}$ (2 V) range at these junction-resistances, one can find departures in $\mathcal{E}(\boldsymbol{r})$ from the average, throughout every unit cell in the FOV. To do so, we first use Fourier filtering to remove the structural supermodulation with wavevector $\boldsymbol{Q}_{SM}$ from measured $g(\boldsymbol{r}, V)$; this is necessary because the supermodulation is known to modulate the charge-transfer energy[25] with an amplitude of approximately 150 meV. We then integrate each measured $g(\boldsymbol{r}, V)$ followed by evaluation of $I_+(\boldsymbol{r}) = \int_0^{V_{max}} g(\boldsymbol{r}, V)dV$, the integrated DOS giving rise to the current for all positive energy states and similarly for the negative energy states, $I_-(\boldsymbol{r}) = \int_{V_{min}}^0 g(\boldsymbol{r}, V)dV$, where $\overline{I_\pm}$ are the equivalent integrals for the spatially averaged spectrum $\overline{g(V)}$. We normalize the current by the junction resistance, which is given by $1/(g_{max} - g_{min})$, doing so leads to the energy variations of the charge transfer band. Both filled and empty states are used in this calculation, as explained in Methods Section C. Next to estimate the variations in the charge transfer energy away from its mean we use

$$\delta'\mathcal{E}(\boldsymbol{r}) \equiv [(\overline{I_+} - I_+(\boldsymbol{r})) - (\overline{I_-} - I_-(\boldsymbol{r}))]/(g_{max} - g_{min}) \qquad (2)$$

which averages over a wide range of tunnel conductances $g$ between $g_{min}$ (0.01 nS) and $g_{max}$ (0.22 nS). This integration algorithm is demonstrated to provide an improved signal-to-noise ratio (Methods Section A & C and Extended Data Fig. 1) in measuring the *variations* of charge transfer energy compared to the previous algorithm[25]. Finally, the Lawler-Fujita procedure[26] is used to morph the simultaneously measured topograph $T'(\boldsymbol{r})$ into a perfectly periodic tetragonal image $T(\boldsymbol{r})$ (Fig. 2a and Methods Section D) in which the geometry of every unit cell is equivalent; the identical transformation carried out on $\delta\mathcal{E}'(\boldsymbol{r})$ yields an image of charge transfer energy variations $\delta\mathcal{E}(\boldsymbol{r})$ that is equally unit-cell periodic[26-29]. This $\delta\mathcal{E}(\boldsymbol{r})$ and its power spectral density Fourier transform $\delta\mathcal{E}(\boldsymbol{q})$ become the basis for studies of intra-unit-cell symmetry breaking, with Fig. 2c showing the configuration of $\delta\mathcal{E}(\boldsymbol{r})$ measured in the FOV of Fig. 2a. The distribution of measured $\delta\mathcal{E}$ values of Fig. 2c is shown in Fig. 2d, and yields an RMS value $\delta\mathcal{E}_{RMS} \approx 90\,\text{meV}$. As demonstrated below, this distribution is dominated by the spatial arrangements of an intra-unit-cell ordered state.

**Intra-Unit Cell Symmetry Breaking of Charge Transfer Energy**

*7*  Next, we explore the symmetry of intra-unit-cell structure of $\delta\mathcal{E}(\boldsymbol{r})$ by studying the values of $\delta\mathcal{E}(\boldsymbol{q})$ measured at the two symmetry-inequivalent Bragg peaks $\boldsymbol{Q}_x^B = 2\pi/a_0 (1,0)$; $\boldsymbol{Q}_y^B = 2\pi/a_0 (0,1)$ of the CuO$_2$ plane [26,27]. Figure 2e is the measured power spectral density Fourier transform of the simultaneous topograph $T(\boldsymbol{q})$ with Fig. 2f showing these data plotted along the two orthogonal axes. The magnitudes of the two Bragg peaks are indistinguishable ($T(\boldsymbol{Q}_x^B) \approx T(\boldsymbol{Q}_y^B)$) within error bars, meaning that neither the crystal nor the STM tip breaks the intra-unit-cell rotational symmetry. In contrast, Fig. 2g shows the measured magnitude of $\delta\mathcal{E}(\boldsymbol{q})$ while Fig. 2h shows these data plotted along the two orthogonal axes. Now the magnitudes of the two Bragg peaks in $\delta\mathcal{E}(\boldsymbol{q})$ are quite different (a repeatable observation see Methods Section E), revealing the existence of $\boldsymbol{Q}$ = 0 rotational symmetry breaking of the electronic structure at the charge transfer energy range in Bi$_2$Sr$_2$CaCu$_2$O$_{8+x}$. To double check, we also demonstrate directly by intra-unit-cell imaging in real space, that this rotational symmetry breaking is a valid empirical property at the charge transfer energy scale of all unprocessed high voltage sublattice resolved $dI/dV(\boldsymbol{r}, eV)$ data (Methods Section E).

**Ising Domains of Orbital Ordering**

*8*  This motivates development of a CuO$_2$ orbital order parameter for efficient visualization of intra-unit-cell energy splitting variations. We focus specifically on the two planar oxygen atom sites within each CuO$_2$ unit cell. First, we sample the values of $\delta\mathcal{E}(\boldsymbol{r})$ surrounding oxygen atom sites along the *x*-axis from Cu atom $\boldsymbol{R}_{ij}$ and label them $\delta\mathcal{E}_{Ox}(\boldsymbol{R}_{ij} - a\hat{\boldsymbol{x}}/2) : \delta\mathcal{E}_{Ox}(\boldsymbol{R}_{ij} + a\hat{\boldsymbol{x}}/2)$; equivalently we sample the two nearest planar oxygen atom sites along the *y*-axis from $\boldsymbol{R}_{ij}$ and label them $\delta\mathcal{E}_{Oy}(\boldsymbol{R}_{ij} - a\hat{\boldsymbol{y}}/2) : \delta\mathcal{E}_{Oy}(\boldsymbol{R}_{ij} +$ $a\hat{\pmb{y}}/2$), as shown inset to Fig. 3c, where $a$ is the CuO$_2$ lattice constant. Using this approach, a nematic order parameter $N_{\mathcal{E}}(\pmb{r})$ may be defined[26,30] as the difference between the average $\delta\mathcal{E}_{Ox}(\pmb{r})$ and $\delta\mathcal{E}_{Oy}(\pmb{r})$ inside each unit cell:

$$N_{\mathcal{E}}\left(\pmb{R}_{ij}\right) = \frac{\left[\delta\mathcal{E}_{Ox}(\pmb{R}_{ij} + a\hat{\pmb{x}}/2) + \delta\mathcal{E}_{Ox}(\pmb{R}_{ij} - a\hat{\pmb{x}}/2)\right]}{2} - \frac{\left[\delta\mathcal{E}_{Oy}(\pmb{R}_{ij} + a\hat{\pmb{y}}/2) + \delta\mathcal{E}_{Oy}(\pmb{R}_{ij} - a\hat{\pmb{y}}/2)\right]}{2}$$

(3)

Figure 3a shows our measured $N_{\mathcal{E}}(\pmb{r})$ from the FOV of Fig. 2a, where the continuous image is achieved by gaussian smoothing of $N_{\mathcal{E}}(\pmb{R}_{ij})$ with radius shown by a small circle in image. Strong breaking of intra-unit-cell C$_4$ symmetry is now observed in $\delta\mathcal{E}(\pmb{r})$, while the predominant disorder is revealed as Ising domains of opposite sign $N_{\varepsilon}$ (a phenomenology that cannot be the result of anisotropy of the scan tip). The relative preponderance of the two orbital order domains is about 2:1 in these $p \approx 0.17$ hole-doped samples (Methods Section E), indicating that the quenched disorder from dopant ions pins the orbital ordered domains so that some domains of the opposite sign to the predominant order are stabilized. This intra-unit-cell structure in charge-transfer energy implies a redistribution of electric charge that breaks rotational symmetry in the form of a charge quadrupole. But each such charge quadrupole has two possible orientations relative to the local environment, with two different energies (Fig. 3b): it is thus an electronic quadrupolar two-level system. Figure 3c shows the histogram of measured $N_{\mathcal{E}}(\pmb{r})$ in Fig. 3a, whose RMS value is approximately 25 meV. The histogram of measured $|N_{\mathcal{E}}(\pmb{r})| < 5$ meV within proposed Ising domain walls is presented in Fig. 3d, representing a dense reservoir of thermally active quadrupolar two-level systems which should undergo rapid thermal fluctuations at finite temperatures. Such two-level systems occupy an area fraction of approximate 25% in this sample (Methods Section F). The dynamics of these two-level systems will be the focus of future studies (Methods Section F). Another issue is the size of the Ising domains, the largest of which (when averaged over all FOV (Methods Section E)) typically has a size of 30 nm$^2$, equivalent to approximately 200 CuO$_2$ unit cells. In the same FOV of these studies we can identify the locations of the interstitial oxygen dopant ions (Methods Section H). When superimposed on the simultaneously measured $N_\varepsilon(\boldsymbol{r})$ images (Fig. 3e), we find a propensity for these ions to lie in the domain walls (Fig. 3f) between the orbitally ordered regions (Methods Section H).

**Determination of Energy Splitting of $O_x$ and $O_y$ Orbitals**

**9**     To epitomize the internal electronic structure of orbitally ordered $CuO_2$, from Fig. 3a we select all regions with $N_\varepsilon > +5$ meV (domain $\alpha$) and all regions with $N_\varepsilon < -5$ meV (domain $\beta$). For each of these two zones, we first average the topographic image $T(\boldsymbol{r})$ over all $CuO_2$ unit cells contained therein (Methods Section F). Figures. 4a, b show that neither the topography nor the scan-tip break the intra-unit-cell $C_4$ symmetry. We then present $\delta\mathcal{E}(\boldsymbol{r})$ unit-cell-averaged over exactly the same two zones. The results, shown in Figs. 4c, d, vividly demonstrate how strongly the intra-unit-cell rotational symmetry is broken at the charge transfer energy, within the two orthogonal Ising nematic domains. Focusing on the Cu sites in Fig. 4c, d reveals breaking of rotational symmetry, while the $O_x$ and $O_y$ sites in Fig. 4c, d exhibit different charge-transfer energies with separation on the 50 meV scale (repeatable observation reported in Methods Section E). Indeed, this intra-unit-cell splitting of the charge transfer energies can be directly seen from unprocessed data where the d$I$/d$V$ spectra are identified separately at the $O_x$ and $O_y$ sites within the same unit cell (Methods Section E). These results are indicative that the $2p^6$ oxygen orbital at $O_x$ is separated from the upper Cu band by a different energy than that at $O_y$, or equivalently the existence of intra-unit-cell orbital order. Many characteristics of $CuO_2$ intra-unit-cell orbital ordering become evident in Figs. 3-4: the spatial structure of $\delta\mathcal{E}(\boldsymbol{r})$ within $CuO_2$ including the specific values at each Cu and inequivalent O site; the energy scale for these orbital ordering phenomena near 50 meV; the statistical distribution $\delta\mathcal{E}(\boldsymbol{r})$ due to the short correlation length Ising domains; and the population of unit-cell localized quadrupolar two-level systems whose energy barriers range continuously from zero. Overall, Figs. 3-4 show that the $\boldsymbol{Q}$ = 0 splitting between $\mathcal{E}(\boldsymbol{r})$ at the two oxygen sites within a $CuO_2$ unit cell is near 50 meV for $Bi_2Sr_2CaCu_2O_{8+x}$ samples with hole-density $p \approx 0.17$, but that long range orbital order is absent resulting in two populations of Ising nematic domains.

**10** These data may be compared to earlier X-ray scattering studies of orbital ordering across the $La_{2-x}Ba_xCuO_4$ materials family. In these crystals, signatures of intra-unit-cell symmetry breaking should occur as anisotropy in the scattering tensor at planar Cu sites. Here, it is the intensity of the (001) scattering peak at the Cu L-edge, which directly measures the electronic nematicity of the Cu $3d$ states within a single $CuO_2$ plane, and provides the indications of intra-unit-cell nematicity[31]. Our Figs. 4c, d also show that $C_4$ symmetry is broken in the charge-transfer electronic structure at the Cu sites (yellow dots) which, if due to Coulomb potentials inside the $CuO_2$ unit cell, also indicates quadrupolar charge distributions surrounding the Cu sites, in agreement with Ref. 31. With respect to other order parameters, we note that the orbitally ordered state at $\boldsymbol{Q}$ = 0 would not have any direct relation to the surface supermodulation or a perfectly harmonic charge density wave (CDW) state at finite $\boldsymbol{Q}_{\text{CDW}} = (2\pi/4a_0, 0); (0, 2\pi/4a_0)$ because any distinctions between $O_x$ and $O_y$ inside the unit-cell due to the density wave, average to 0 over one period[26,27]. However, higher harmonics of a CDW can, in principle, produce composite order parameters which break symmetry at $\boldsymbol{Q}$ = 0, as can "vestigial" CDW states in the presence of disorder[32]. Another important subject is the effect of electron-lattice interactions: while the predominant theoretical motivation for such a state has been based on the repulsive $V_{pp}$ potential between the oxygen atoms[2-9] (Fig. 1b), it is inevitable that such orbital order will also couple to the $B_{1g}$ phonon mode. If this mode softens in $Bi_2Sr_2CaCu_2O_{8+x}$, the O atom elevated above the $CuO_2$ plane experiences a different potential than its neighbor which is depressed below it[33]. One surprising new revelation is the distribution of spatially localized, low activation energy two-level systems (TLS) within the boundaries between orbital ordered domains. These are the electronic quadrupoles within each unit cell (Fig. 4c, d) whose energy barrier between the two orientations of $N_\varepsilon$ tends to zero (Fig. 3d) making thermally activated fluctuations conceivable. Ultimately, as to the key issue of the 50 meV scale of energy splitting between $O_x:O_y$, its precise microscopic cause is an outstanding theoretical challenge apparently requiring realistically parameterized theory for the three-orbital model (Eqn. 3) within the $CuO_2$ unit cell.

**11**     Nevertheless, significant consequences do emerge from these studies. First, when physically realistic models of $CuO_2$ electronic structure are used (e.g. Eqn. 1 with parameters relevant to real materials), Coulomb interactions between electrons at the two crystal-equivalent oxygen sites should generate intra-unit-cell rotational symmetry breaking i.e. orbital ordering[2-9]. By introducing sublattice resolved $\mathcal{E}(\boldsymbol{r})$ imaging techniques to $CuO_2$ studies, this state has now been detected (Fig. 4). By visualizing an oxygen-site-specific nematic order parameter $N_\mathcal{E}(\boldsymbol{r})$ we reveal robust domains of $C_2$ symmetry and with typical energy splitting $\approx$ 50 meV between the two intra-unit-cell oxygen sites (Fig. 3). Hence, a strong orbitally ordered state occurs in underdoped $Bi_2Sr_2CaCu_2O_{8+x}$ at the charge-transfer energy scale. Spatially, the state is arranged in Ising domains, which are highly disordered. This disorder appears to be bounded in shape and size by randomly sited oxygen dopant-ions. Moreover, within domain walls as $N_\mathcal{E} \to 0$ (Fig. 3d) we find an ensemble of low energy-barrier electronic quadrupolar two-level systems. Most fundamentally, the microscopic mechanism proposed theoretically[2-9] for the cuprate nematic phase i.e. orbital order between oxygen orbitals at the two separate oxygen sites of $CuO_2$, is highly consistent with the observed intra-unit cell rotational symmetry breaking of $\mathcal{E}(\boldsymbol{r})$ which splits the energy between the two oxygen atoms by $\sim$ 50 meV in $Bi_2Sr_2CaCu_2O_{8+x}$.

**Acknowledgements**: We acknowledge and thank P. Bourges, D. Bounoua, V. Madhavan, and P.Z. Mai for helpful discussions. K.F. and J.C.S.D. acknowledge support from the Moore Foundation's EPiQS Initiative through Grant GBMF9457. S.W. and J.C.S.D. acknowledge support from the European Research Council (ERC) under Award DLV-788932. J.C.S.D. acknowledges support from the Royal Society under Award R64897. N.K., J.C.S.D. and S.O'M acknowledge support from Science Foundation of Ireland under Award SFI 17/RP/5445. S.W. acknowledges support from the John Fell Fund at University of Oxford under the project 0010827. K.F. acknowledges support from the U.S. Department of Energy, Office of Basic Energy Sciences, under contract number DEAC02-98CH10886. H.E. acknowledges support from JSPS KAKENHI (No. JP19H05823). P.D.J. acknowledges support by QuantEmX grant GBMF9616 from ICAM / Moore Foundation and by a Visiting Fellowship at Wadham College, Oxford, UK. This research program was also advanced through support from the National Science Foundation under Grant No. NSF PHY-1748958 at the Kavli Institute for Theoretical Physics of U.C.S.B.; S.W. and J.C.S.D. thank KITP for its hospitality.

**Author Contributions:** S.W., N.K. and J.C.S.D. conceived the project. H.E. and S.U. synthesized and characterized the samples; K.F. and J.C.S.D. carried out experimental measurements; S.W., N.K., K.F. and S.O'M developed and carried out the comprehensive analysis. P.D.J, J.C.S.D and S.O'M supervised the research and wrote the paper with key contributions from S.W. and N.K. The manuscript reflects the contributions and ideas of all authors.

**Author Information** Correspondence and requests for materials should be addressed to J.C. Seamus Davis jcseamusdavis@gmail.com or Shuqiu Wang shuqiucwang@gmail.com.

Figure 1

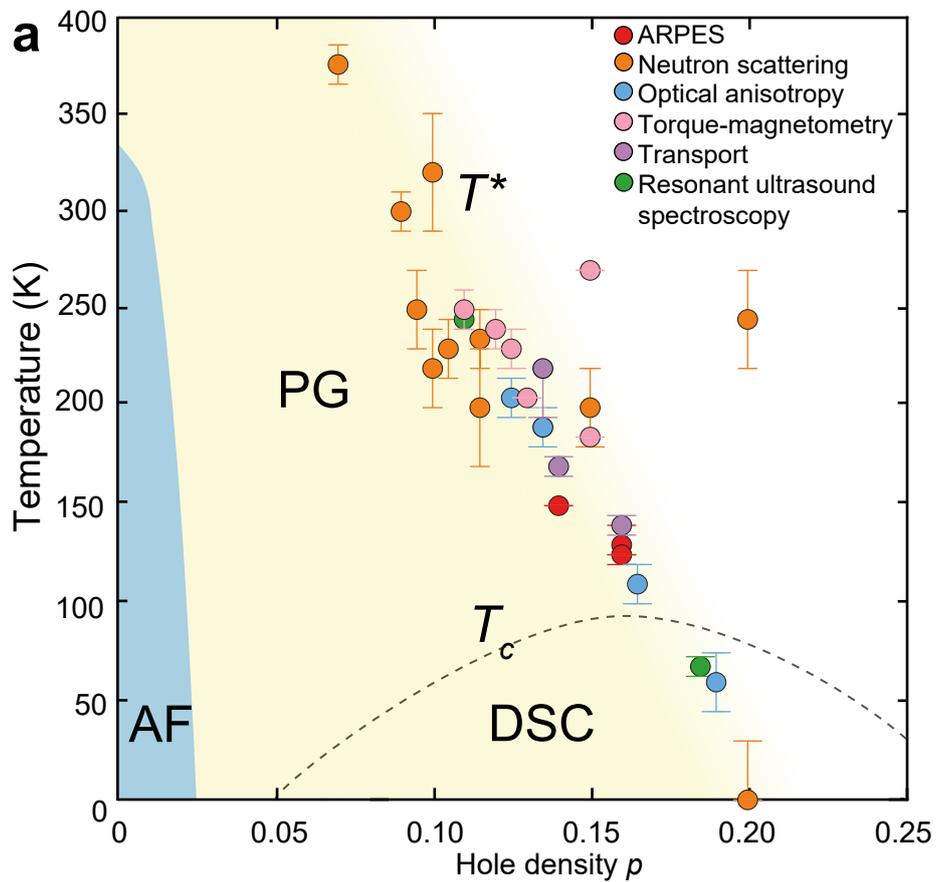

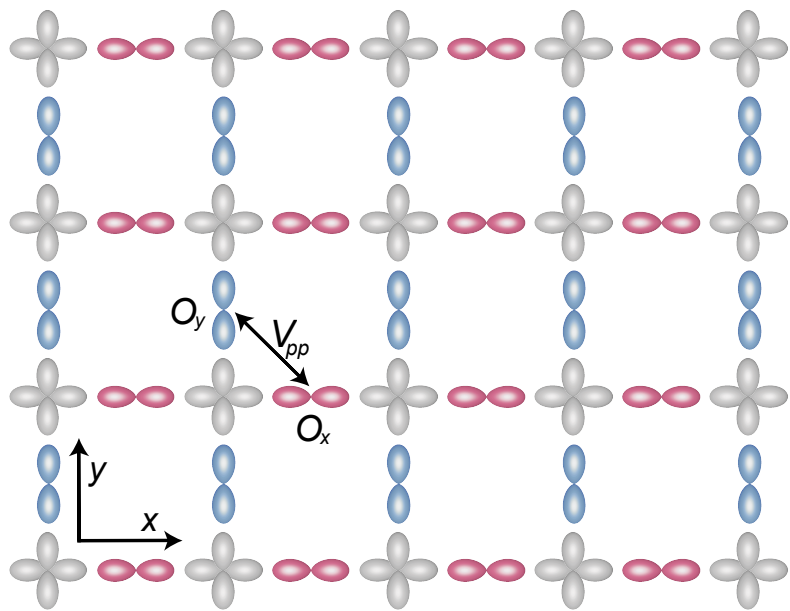

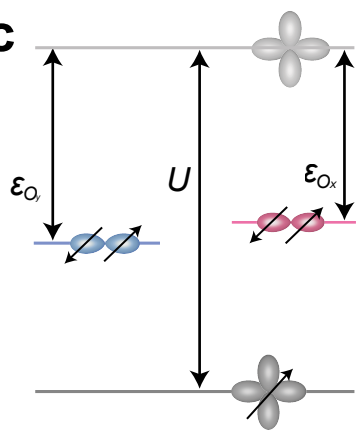 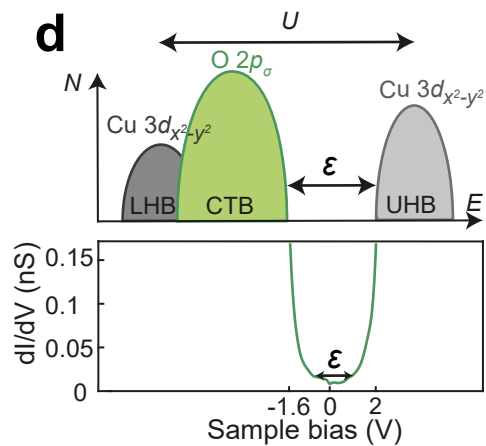

**Fig. 1: Concept of CuO$_2$ Intra-unit-cell Orbital Order**

(a) Signatures of nematic phase appearing at the cuprate pseudogap temperature $T^*(p)$.
(b) Schematic of the relevant orbitals in the CuO$_2$ plane depicting the crucial inter-oxygen-site Coulomb interaction $V_{pp}$.
(c) The degeneracy of the Cu $3d_{x^2-y^2}$ orbital (gray) is lifted by the Coulomb energy $U$. The $p_x$ orbital of the oxygen along the *x*-axis of the Cu atom (red) is separated from the upper Cu band by the charge transfer energy $\mathcal{E}_x$. At the oxygen site along the *y*-axis of the Cu atom (blue), its $p_y$ orbital is separated from the upper Cu band by the charge transfer energy $\mathcal{E}_y$.
(d) Top panel shows a schematic density of electronic states where the Coulomb energy $U$ and charge transfer energy $\mathcal{E}$ are indicated. Bottom panel shows a typical measured differential conductance $g(V)$ spectrum where the top of the lower band and the bottom of the upper band are visualized, their separation being a direct measure of charge transfer energy $\mathcal{E}$[25].

Figure 2

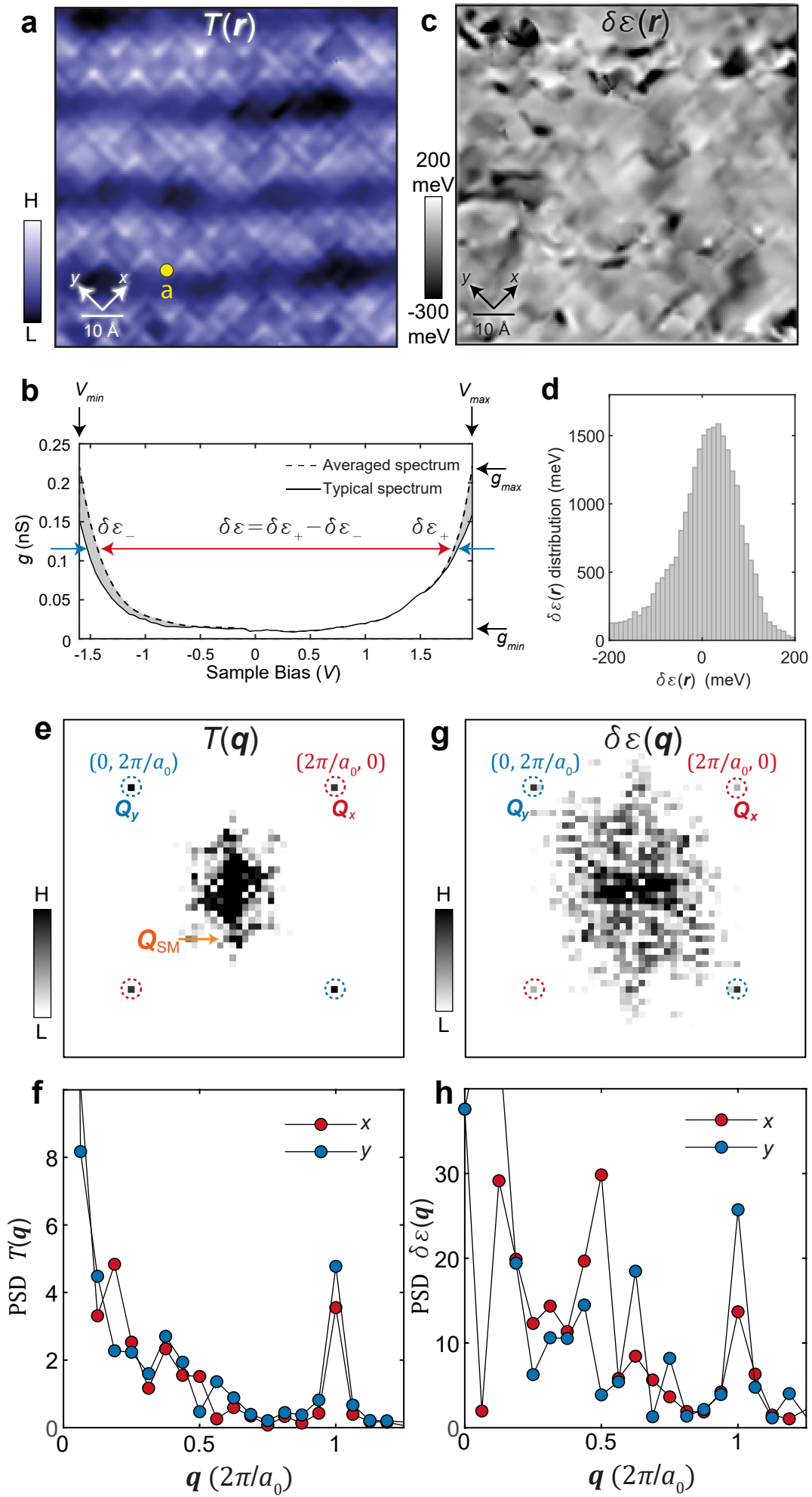

**Fig. 2**: **Visualizing CuO$_2$ Charge-transfer Energy Variations $\delta\mathcal{E}(r)$**

(a) Topographic image of BiO surface of Bi$_2$Sr$_2$CaCu$_2$O$_{8+x}$ studied here. Setpoint for topography was -750 mV at 25 pA. The bulk incommensurate crystal supermodulation is seen clearly in Fig. 2a. It is at 45° to, and therefore is the mirror plane between, the x- and y-axes as always[26]. The orbital ordering domains are not influenced by supermodulation for this symmetry reason. Furthermore, we compare three independent orbital ordering domains analyzed with and without the supermodulation in Methods Section J. Despite crystal structure strongly modulated by the supermodulation, the orbital ordering domains remain consistent. Therefore the supermodulation has no discernable influence on the intra-unit-cell symmetry breaking of $\delta\mathcal{E}(r)$ discussed in this paper.

(b) Typical high-voltage differential conductance spectra $g(r, V)$ from (a) show as a solid black curve while the spatially averaged spectrum $\overline{g(V)}$ is shown as a dashed curve. The exemplary spectrum is measured at location a denoted as a yellow dot in (a). Such high junction resistances of 85 GΩ or large tip-sample distances preclude effects on $g(V)$ of the tip-sample electric field. Separation between the lower band and the upper band is clearly very distinct for the exemplary spectrum (blue arrows) and for the average spectrum (red double-headed arrow). We estimate that energy difference $\delta\mathcal{E}$ using Eqn. 2. Setpoint for the differential conductance map $g(r, V)$ was 600 mV at 7 pA.

(c) Typical visualization of charge-transfer energy variations map $\delta\mathcal{E}(r)$ from (a).

(d) Histogram of charge-transfer energy variations $\delta\mathcal{E}$ in (c).

(e) Power spectral density Fourier transform $T(q)$ of the topographic image measured simultaneously as (c). The $Q_{SM}$ peaks (orange arrow) are signal from the supermodulation which is a quasi-periodic modulation along the (1,1) direction.

(f) The linecuts from $q = (0, 0)$ to $(1, 0)2\pi/a_0$ and from $q = (0, 0)$ to $(0, 1)2\pi/a_0$ in $T(q)$ from (e). Linecuts of $T(q)$ show virtually indistinguishable values at $Q_x = (1, 0)2\pi/a_0$ and $Q_y = (0, 1)2\pi/a_0$ thus the power spectral density $T(q)$ does not break C$_4$ symmetry at its Bragg peaks.

(g) Power spectral density Fourier transform $\delta\mathcal{E}(q)$ of charge-transfer energy map from (c). The $Q_{SM}$ peaks are removed in $\delta\mathcal{E}(q)$.

(h) The linecuts from $q = (0, 0)$ to $(1, 0)2\pi/a_0$ and from $q = (0, 0)$ to $(0, 1)2\pi/a_0$ in the $\delta\mathcal{E}(q)$ from (g). The power spectral density $\delta\mathcal{E}(q)$ breaks C$_4$ symmetry at its Bragg peaks because the plots of $\delta\mathcal{E}(q)$ show anisotropy at $Q_x = (1, 0)2\pi/a_0$ and $Q_y = (0, 1)2\pi/a_0$. This is the first direct evidence of intra-unit-cell rotational symmetry breaking at the charge transfer energy scale in cuprates.

Figure 3

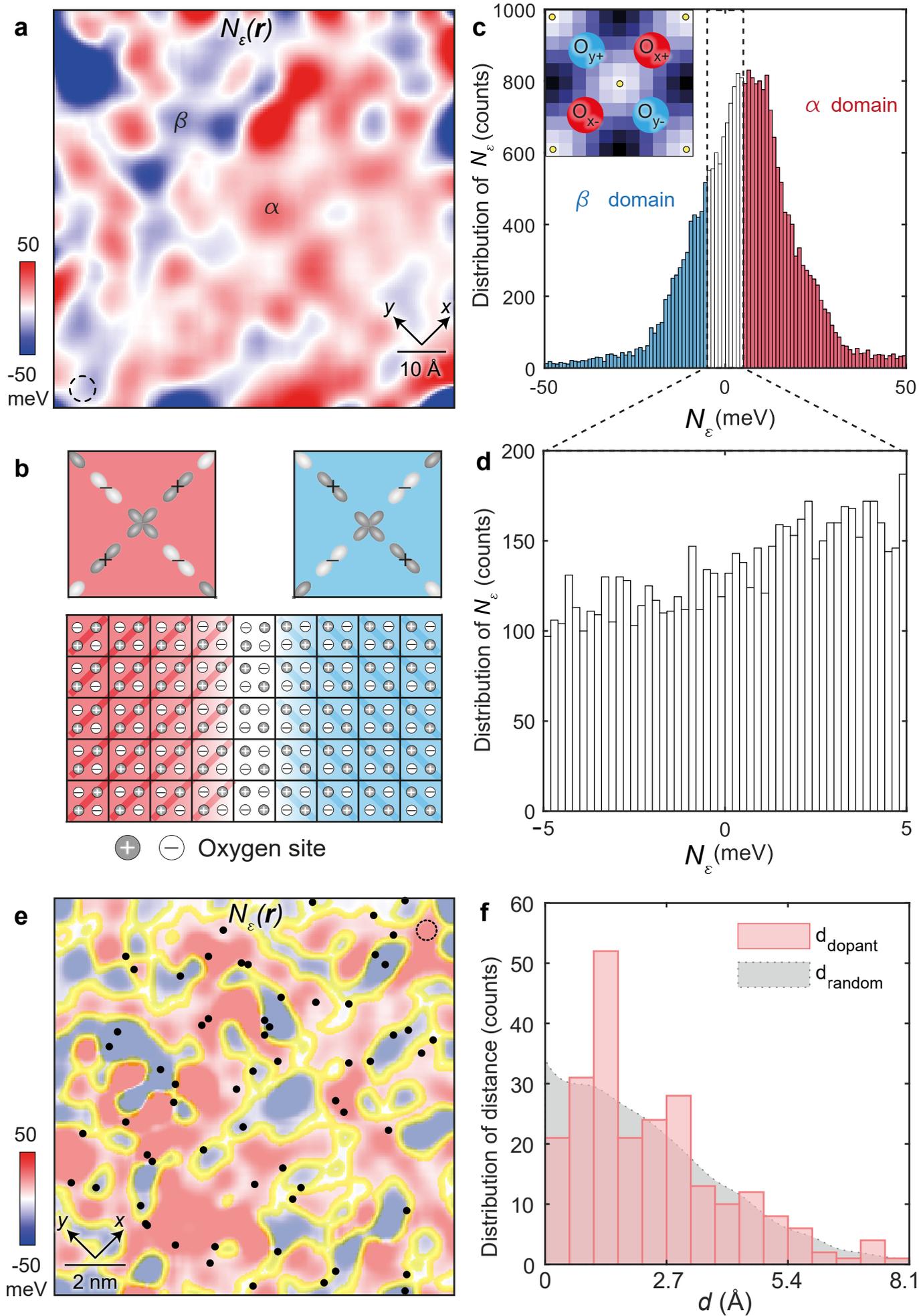

**Fig. 3: Intra-unit-cell d-symmetry oxygen-specific characteristics of charge-transfer energy**

(a) Image of nematic intra-unit-cell order parameter $N_{\varepsilon}(R_{i,j})$ sampled on oxygen sites. $\delta\varepsilon_{O_x(O_y)}$ represent the four oxygen sites of each unit cell (Inset of Fig. 3c) from the FOV of Fig. 2a. The continuous $N_{\varepsilon}(r)$ image is achieved by gaussian smoothing of $N_{\varepsilon}(R_{ij})$ with a radius of 3.2 Å shown by a small circle in image. Powerful breaking of intra-unit-cell $C_4$ symmetry is now clearly observed in $\delta\varepsilon(r)$, while the predominant disorder in revealed as Ising domains of opposite sign $N_{\varepsilon}$. We may also use a d-symmetry measure $D_{\varepsilon}(r)$ that segregates the whole $CuO_2$ unit cell into four quadrants (Methods Section F). These images of $D_{\varepsilon}(r)$ are virtually identical to $N_{\varepsilon}(r)$ with typical correlation between them $N_{\varepsilon}(r)$: $D_{\varepsilon}(r)$ = 0.93.

(b) Schematic of charge-quadrupole two-level systems within Ising domain walls. Top panel schematizes the charge distribution in the four oxygen sites of one $CuO_2$ unit cell consequent to the intra-unit-cell charge transfer symmetry breaking. Bottom panel shows a schematic of two orbitally ordered Ising domains consisting of orthogonally oriented charge quadrupoles, and a domain wall within which the energy barrier to fluctuations in the charge quadrupolar orientation can be exceeded at finite temperatures. The oxygen sites with higher charge transfer energy $\varepsilon$ are indicated by darker circles and the oxygen sites with lower $\varepsilon$ are indicated by lighter circles.

(c) Histogram of measured $N_{\varepsilon}(r)$ in (a), whose RMS value is approximately 25 meV. This histogram represents a combination of two populations of ordered Ising nematic domains plus all the randomized values of $\delta\varepsilon(r)$ within the intervening domains walls. The inset shows the unit cell sampled on four oxygen sites (oxygen atoms along the x-axis in red and along the y-axis in blue), from which the nematic order parameter is calculated (Methods Section G). The location of the Cu sites is indicated by yellow circles.

(d) Histogram $N_{\varepsilon}$ from all non-ordered regions of (a). Such regions are identified as the white domain walls in (A) where $|N_{\varepsilon}| < 5$ meV. The thermal energy of 5 meV is equivalent to approximately 60 K. The cut-off of 5 meV is validated by estimating the width of domain boundaries using the correlation length (Methods Section F).

(e) The location of oxygen dopant ions (black dots) measured simultaneously as $N_{\varepsilon}(r)$ (Methods Section H). They occur proximate to the domain walls where $|N_{\varepsilon}(r)| < 5$ meV (domain walls are highlighted in yellow contour). The area of the dopant atoms only accounts for approximately 10% of the total FOV thus dopants do not affect $\delta\varepsilon(r)$ measurements.

(f) Histogram of the shortest distance $d_{\text{dopant}}$ between the dopant ion and a domain wall (pink bars). Histogram of the expected averaged distance $d_{\text{random}}$ between the simulated Poisson random point and its nearest point in the domain wall (grey area), which has no correlation with $N_{\varepsilon}(r)$; (Methods Section H).

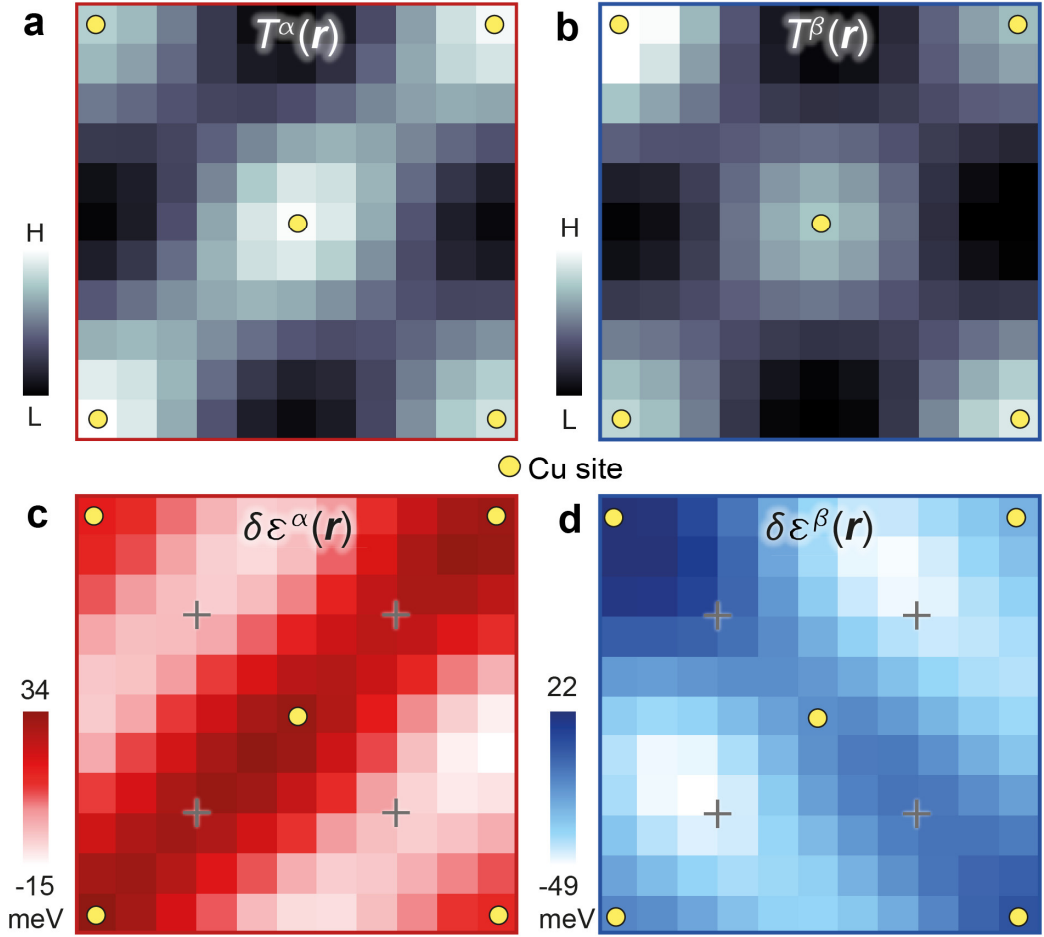

**Fig. 4: Sublattice-resolved charge-transfer energy and orbital ordering**

(a) Unit-cell averaged structure of $T(r)$ averaged over all regions in Fig. 3a where $N_\varepsilon > +5$ meV.

(b) Unit-cell averaged structure of $T(r)$ averaged over all regions in Fig. 3a where $N_\varepsilon < -5$ meV. In both (a) and (b) the $C_4$ symmetry is preserved.

(c) Unit-cell averaged structure of $\delta\mathcal{E}(r)$ averaged over all regions in Fig. 3a where $N_\varepsilon > +5$ meV. The charge transfer energy strongly breaks $C_4$ rotation symmetry about every Cu site (yellow dots) and consequently there is an energy splitting of approximately 50 meV between the charge transfer energies at the two crystal-equivalent oxygen sites (indicated by the crosses). Unprocessed high voltage d$I$/d$V$ images reveal directly the subtending data (Methods Section E).

(d) Unit-cell averaged structure of $\delta\mathcal{E}(r)$ averaged over all regions in Fig. 3a where $N_\varepsilon < -5$ meV. Virtually identical phenomena as in (c) but rotated by 90 degrees. The difference between (c) and (d) is the difference in the internal structure of the CuO$_2$ unit cell in the two distinct Ising domains of orbital order. Unprocessed high voltage d$I$/d$V$ images reveal directly the subtending data (Methods Section E).

Methods for

**Discovery of Orbital Ordering in $Bi_2Sr_2CaCu_2O_{8+x}$**

Shuqiu Wang, Niall Kennedy, K. Fujita, S. Uchida, H. Eisaki,

P.D. Johnson, J.C. Séamus Davis and S.M. O'Mahony

Correspondence to: J.C. Seamus Davis jcseamusdavis@gmail.com
or Shuqiu Wang shuqiucwang@gmail.com

**Methods**

The high-voltage spectroscopic imaging scanning tunneling microscope (SISTM) measurements of $Bi_2Sr_2CaCu_2O_{8+x}$ crystals are carried out in a custom-built SISTM. The sample is grown using the floating zone method with doping controlled by oxygen depletion. The hole doping level studied in this paper is $p\sim17\%$. The samples are cleaved in a cryogenic ultrahigh vacuum at a temperature of ~4.2 K so that the atomically flat BiO termination is revealed. The samples are then immediately inserted into the STM head. All measurements are carried out using tungsten tips at a base temperature of ~4.2 K.

**A. Visualization of Spatially Modulating Charge Transfer Energy**

The spatially resolved charge transfer energy $\mathcal{E}(\boldsymbol{r})$ may be visualized in cuprates using high voltage[1–3] differential tunnel conductance $g(\boldsymbol{r},V) \equiv dI/dV(\boldsymbol{r},V)$ imaging of $Bi_2Sr_2CaCu_2O_{8+x}$[4]. Here, $g(\boldsymbol{r},V)$ was visualized in the range $-1.6\text{ V} \leq V \leq 2\text{ V}$ and at the very high junction resistance $R_N \approx 85\text{ G}\Omega$ ($V_S = 600$ mV: $I_S = 7$ pA) necessary to suppress tip-induced electric field effects. Fig. 2b in the main text shows a typical example of a $g(\boldsymbol{r},V)$ spectrum measured at $T = 4.2$ K[4]. The edges of the filled lower band and the empty upper band can be identified from the appearance of extremely rapid increase in the density of states. Here the value of the charge transfer energy $\mathcal{E}(\boldsymbol{r})$ is estimated at every location by subtracting these band edges at a constant differential conductance $G \approx 20$ pS as follows:

$$\mathcal{E}(\boldsymbol{r}) = V_{\text{empty}}(G,\boldsymbol{r}) - V_{\text{filled}}(G,\boldsymbol{r}) \qquad (M1)$$

Direct visualization of the charge transfer energy has been reported from several STM groups. McElroy *et al.* discussed the difficulty of high voltage mapping on $Bi_2Sr_2CaCu_2O_{8+x}$ in detail and demonstrated how this can be overcome[1]. Cai *et al.* successfully sampled the charge transfer gap in both La-Bi2201 (1.7 eV) and $Ca_{2-x}Na_xCuO_2Cl_2$ (2.2 eV) but a full map spanning *r*-space was not reported[3]. Ruan *et al.* studied both single- and double-layer families of $Ca_{2-x}Na_xCuO_2Cl_2$ and Bi-2201 and Bi-2212 using 77 K STM, again successfully determining the charge transfer gap (2 eV, 1.4 eV, 1.5 eV and 1.0 eV respectively) but only point spectra were measured[5]. In Wang *et al.* there is a charge transfer energy STM study of Bi-2201, Bi-2212 and Bi-2223 finding 1.5 eV, 1.0 eV and 0.7 eV respectively, but again only point spectra are reported[6]. By using STM the charge transfer energy values are found in excellent agreement with optical spectroscopy and ARPES measurements[7]. Therefore, charge transfer measurements are well established by the STM community as the separation in the dI/dV spectra between the upper edge of the charge transfer band and the lower edge of the upper Hubbard band.

Although charge transfer energy measurements have been successful in STM studies, their real space distribution and especially atomic resolution have been challenging to achieve. The difficulty here is maintaining a stable tunneling junction at high bias requires an unusually high junction resistance. In this paper, we achieve intra unit cell resolution in the charge transfer energy at such high junction resistances and thus resolve variations of the charge transfer energy on the $O_x$ and $O_y$ sites.

The crystal supermodulation is a quasi-periodic lattice modulation along the (1,1) direction in the crystal structure at the wavevector $\boldsymbol{Q}_{SM}$. The resulting $\mathcal{E}(\boldsymbol{r})$ map for $Bi_2Sr_2CaCu_2O_{8+x}$ shows strong modulations at the same wavevector, with an amplitude of $\approx 0.3$ eV and has a spatial average value of $\langle\mathcal{E}(\boldsymbol{r})\rangle \approx 1.2$ eV which agrees well with the charge-transfer energy of $Bi_2Sr_2CaCu_2O_{8+x}$ as measured independently using a variety of different experimental techniques[5,7,8]. Analysis of the supermodulation influence on $\mathcal{E}(\boldsymbol{r})$ and the corresponding supermodulation influence on the electron-pair density $n_p(\boldsymbol{r})$ measured by scanned Josephson tunnelling microscopy have revealed that the charge-transfer energy $\mathcal{E}(\boldsymbol{r})$ controls the density of condensed electron-pairs $n_p(\boldsymbol{r})$ as $d\bar{n}_P/d\mathcal{E} \approx -0.81 \pm 0.17$ eV$^{-1}$ in Ref. 4. This is in quantitative agreement with cluster dynamical mean field theory (cDMFT) solutions of the Emery three-band model which predict $d\bar{n}_P/d\mathcal{E} = 0.93 \pm 0.1$ eV$^{-1}$ for $Bi_2Sr_2CaCu_2O_{8+x}$[9]. The quantitative agreement between theory and experiment here serves a dual purpose. Firstly, it provides strong evidence that charge-transfer superexchange is the electron-pairing mechanism of the cuprate high temperature superconductors and secondly, it provides further justification of the experimental $\mathcal{E}(\boldsymbol{r})$ visualization technique described above and used throughout this study.

Hence, Extended Data Fig. 1 compares the images of charge transfer energy $\mathcal{E}(\boldsymbol{r})$ derived from the simple algorithm in Eqn. M1 and the more general algorithm for charge transfer energy variation $\delta\mathcal{E}(\boldsymbol{r})$ used in the main text. The images of $\mathcal{E}$ and $\delta\mathcal{E}$ bear high similarities in real space and reciprocal space. Their linecuts show identical and strong symmetry breaking at the Bragg peaks where the ratio of the intensity of the Bragg peak in the *y*-direction to the Bragg peak in the *x*-direction reveals, $\frac{Q_y}{Q_x} = 1.9$. Thus, our general algorithm calculating the charge transfer energy variation $\delta\mathcal{E}(\boldsymbol{r})$ throughout this paper is consistent with the earlier simple algorithm[4] for calculating $\mathcal{E}(\boldsymbol{r})$. Furthermore, Extended Data Figs. 1 d, e show the histograms of the distribution of the charge transfer energy $\mathcal{E}(\boldsymbol{r})$ and variation of the charge transfer energy $\delta\mathcal{E}(\boldsymbol{r})$ respectively. Clearly, the $\delta\mathcal{E}(\boldsymbol{r})$ histogram features a narrower σ and therefore features a higher signal-to-noise ratio. Thus, we implement our algorithm to measure charge transfer energy variation $\delta\mathcal{E}(\boldsymbol{r})$ throughout our paper in order to utilize its enhanced signal-to-noise ratio.

**B. Experimental evidence of $\boldsymbol{Q}$=0 rotational symmetry breaking in cuprates.**

The existence of a nematic state in cuprates, which breaks rotational symmetry at $\boldsymbol{Q} = 0$ has been demonstrated experimentally. This state has been widely reported based on multiple techniques, (Main Text Fig. 1a). Neutron scattering (NS) is the most widely used[10-15], here polarized neutrons are used to detect the anisotropy at the Bragg peaks, indicating IUC symmetry breaking. Since translational symmetry is found to be preserved, this can be explained by symmetry breaking inside each unit cell. Other experimental methods used are angle-resolved photoemission spectroscopy (ARPES)[16-18], where above $T^*$ left-circularly polarized (LCP) and right-circularly polarized (RCP) light have the same intensity at the mirror plane, but when cooled below $T^*$, a difference in LCP and RCP intensity at the mirror plane emerges, which is a signature of a $\boldsymbol{Q} = 0$ rotational symmetry breaking state. Torque-magnetometry (MT) studies[19,20] have reported a kink of the in-plane anisotropy of the susceptibility at $T^*$ which relates to rotational symmetry breaking. Optical anisotropy (ORA)[21] measurements show a significant change in the second-harmonic of the optical response near $T^*$, whereas the linear response remains unchanged in the same region, this could be explained if bulk inversion symmetry was broken, alongside this nematicity is also seen. In polarization resolved Raman scattering[22] suppression of the susceptibility near $T^*$ is observed. Elastoresistance (Transport) measurements[23] of in-plane anisotropy which onsets near $T^*$ also indicates a nematic state. Resonant ultrasound spectroscopy (RUS) [24] finds a phase transition at $T^*$ by noting discontinuities in the frequencies and widths of the vibrational normal modes of a crystal. STM experiments[25] have detected the intra-unit-cell rotational symmetry breaking representing nematicity in the low energy density of states, and this nematicity exhibits Ising domains which diminish in size and intensity approaching the pseudogap endpoint. In $La_{1.875}Ba_{0.125}CuO_4$[26,27]; $La_{1.65}Eu_{0.2}Sr_{0.15}CuO_4$[26,28]; $La_{1.6-x}Nd_{0.4}Sr_xCuO_4$[26,28,29] resonant soft-x-ray scattering also reveals a nematic phase. This electronic phase transition is explored using resonant x-ray scattering, particularly measuring the [001] Bragg peak. The intensity of the Bragg peak is proportional to the electronic nematic order parameter that is associated with the symmetry breaking of the Cu d-state. The photon energies are sensitive to different atoms inside the cuprate unit cell. When the photon energies sensitive to Cu are used, the emerging [001] intensity reveals a nematic phase. Importantly, the nematic phase is reported across numerous cuprate families, including $YBa_2Cu_3O_{7-x}$[10,11,14,15,19,21,24], $Bi_2Sr_2CaCu_2O_{8+x}$[16,18,22,23,25], $Bi_2Sr_2CuO_{6+x}$[17], $Bi_{2-z}Pb_zSr_{2-y}La_yCuO_{6+x}$[30], $HgBa_2CuO_{4+\delta}$[12,13,20] and $La_{1.875}Ba_{0.125}CuO_4$ [26-29]. But its microscopic mechanism was unknown.

This challenge provides the motivation to search for a charge transfer energy splitting due between the $O_x$ and $O_y$ site due to orbital ordering as predicted, for example, in Kivelson *et al* at the strong coupling limit, three-band Emery model produces a nematic phase [31] or Fischer *et al* study the charge transfer energy using the three-orbital model for the cuprates[32]; or Hardy *et al* report a nematic metal phase which arises from the spontaneous partial orbital polarization of the multiorbital non-Fermi liquid[33]. These orbital ordering theories provide the direct motivation for us to search in real space for cuprate orbital ordering in $Bi_2Sr_2CaCu_2O_{8+x}$ by using STM.

### C. Sublattice-resolved Charge Transfer Energy Visualization

The technical details of measuring the variation of the charge transfer energy $\delta\mathcal{E}(\boldsymbol{r})$ are covered in this section. The averaged differential conductance $g(V)$ is shown as a dashed curve in Extended Data Fig. 2. The variations of the charge transfer energy are determined from the deviation of a point spectrum from the averaged spectrum. The states on the *V*>0 side have a positive integral $I_+(\boldsymbol{r}) = \int_0^{V_{max}} g(V,\boldsymbol{r})\,dV$, where the maximum energy is $V_{max} = 2$ V. The states on the *V*<0 side have a negative integral $I_-(\boldsymbol{r}) = \int_{V_{min}}^{0} g(V,\boldsymbol{r})$ where the minimum is $V_{min} = -1.6$ V. The variation in each integral $I_+(\boldsymbol{r}); I_-(\boldsymbol{r})$ from the average values, $\overline{I_+}; \overline{I_\pm}$, occurs due to the variation in energy separation $\delta\mathcal{E}(\boldsymbol{r})$ between the lower band and the upper band from its average value. To evaluate the energy splitting between the point spectrum and the averaged spectrum, the integral difference is normalized by the difference between the maximum differential conductance $g_{max}$ and the minimum difference conductance $g_{min}$. Here $g_{max}$ = 0.22 nS is given by the maximum value of the field of view averaged spectrum and $g_{min}$ = 0.009 nS is the minimum differential conductance of the averaged spectrum.

Hence, we define
$$(g_{max} - g_{min})\delta\mathcal{E}(\boldsymbol{r}) = [\overline{I_+} - I_+(\boldsymbol{r})] - [\overline{I_-} - I_-(\boldsymbol{r})] \tag{M2}$$
There are two typical cases shown in Extended Data Fig. 2. In Case 1 the separation in energy between lower and upper bands in the point spectrum is less than that of the total field of view averaged. The area integrated below the averaged spectrum is larger than the area below the point spectrum thereby the energy variation is positive $\delta\mathcal{E}(\boldsymbol{r})$>0. In Case 2 the separation in energy between lower and upper bands in the point spectrum is above that of the averaged spectrum. The area integrated below the averaged spectrum is smaller than the area below the point spectrum and this case gives rise to a negative energy variation $\delta\mathcal{E}(\boldsymbol{r})$<0.

The choice of the integration range of conductances in Eq. M2 does not alter the conclusion that $C_4$ symmetry is broken in $\delta\mathcal{E}$. Extended Data Fig. 3 shows $\delta\mathcal{E}(\boldsymbol{r})$, $\delta\mathcal{E}(\boldsymbol{q})$ and the value of the Bragg peaks integrated up to various $g_{max}$ values. The first column in Extended Data Fig.

3 show $\delta\mathcal{E}(r)$ for $g_{max}$ = 0.05 nS, 0.10 nS, 0.15 nS, 0.20 nS, 0.22 nS, respectively. Clearly, the calculations at several $g_{max}$ give rise to highly similar spatial distributions of $\delta\mathcal{E}(r)$. Furthermore, $\delta\mathcal{E}(r)$ spans the same range, -300 meV to 200 meV, independent of the choice of $g_{max}$. Next, the second column in Extended Data Fig. 3 show the Fourier transform of $\delta\mathcal{E}(r)$, $\delta\mathcal{E}(q)$ for the same range of $g_{max}$. All images here show strong disorder at $q$ = 0, with this disorder increasing as $g_{max}$ increases. Also, the intensities of all $Q_y$ Bragg peaks are greater than the intensities of $Q_x$ meaning IUC rotational symmetry is broken in all images. The third column in Extended Data Fig. 3 are linecuts from $q = (0,0)$ to $(1,0)2\pi/a_0$ and from $= (0,0)$ to $(0,1)2\pi/a_0$ in the measured $\delta\mathcal{E}(q)$ images. The ratio of these linecuts at $q = 2\pi/a_0$ is a measure of nematicity. When $g_{max}$ = 0.05 nS, 0.10 nS, 0.15 nS, 0.20 nS, 0.22 nS, the ratios of the transverse averaged intensities of the Bragg peaks are $Q_y/Q_x$ = 1.64, 1.72, 1.79, 1.71, 1.45 respectively. Therefore, the charge transfer energy variations are anisotropic and this IUC rotational symmetry breaking of $\delta\mathcal{E}(r)$ is virtually invariant to the value of $g_{max}$. The wavevector of the supermodulation $Q_{SM}$ is removed so that the intra-unit-cell modulation is more visible.

Extended Data Fig. 4 shows a comparison between $\delta\mathcal{E}(r)$ for the empty states $\delta\mathcal{E}_+(r)$, the charge transfer energy variations determined from $(g_{max} - g_{min})\delta\mathcal{E}_+(r) = [\overline{I}_+ - I_+(r)]$, and for the filled states $\delta\mathcal{E}_-(r)$ determined from $(g_{max} - g_{min})\delta\mathcal{E}_-(r) = [\overline{I}_- - I_-(r)]$ The fluctuations of the upper Hubbard band (UHB), $\delta\mathcal{E}_+(r)$, in Extended Data Fig. 4A is dominated by disorder. The Bragg peaks in the Fourier transform $\delta\mathcal{E}_+(q)$ have almost identical intensity that $Q_y/Q_x$ = 0.98 (Extended Data Fig. 4b). The $\delta\mathcal{E}_-(r)$ in Extended Data Fig. 4d represents the modulations of the edges of the filled lower band. $\delta\mathcal{E}_-(r)$ has no spatial correlation with $\delta\mathcal{E}_+(r)$. The Fourier transform of the filled states $\delta\mathcal{E}_-(q)$ (Extended Data Fig. 4e) displays strong anisotropy in the Bragg peaks with $Q_y/Q_x$ = 1.77 and thus strong nematicity in the charge transfer energy at its lower band edge.

### D. Lawler-Fujita Symmetrization of Charge Transfer Energy Images

Here we describe how to process the image in removing a pico-meter distortion due to the piezo-electronic drift of the STM tip scanner. First, we apply the affine transformations to the topograph, $T'(r)$ and the simultaneously taken electronic structure images, such that the two sets of Bragg wavevectors of Bi atoms in BiO plane, $Q_1$ and $Q_2$ in $T'(r)$ satisfy $|Q_1| = |Q_2|$ and $Q_1 \cdot Q_2 = 0$ after the transformation. Then, we apply the Lawler-Fujita (LF) algorithm[34] to remove the STM tip scanner drift in the data, in which the following steps are taken. The $T'(r)$, in which the STM tip drift is embedded, is given by

$$T'(r) = |A_{Q_1}(r)|\cos\left(Q_1 \cdot (r + u(r))\right) + |A_{Q_2}(r)|\cos\left(Q_2 \cdot (r + u(r))\right), \quad \text{(M3)}$$

where $|A_Q(r)|$ is an amplitude of the complex field $A_Q(r)$ for the wavevector $Q$ and $u(r)$ is the displacement field associated with the STM tip scanner drift that distorts the $T'(r)$[34]. To get the displacement field $u(r)$ for a periodic modulation at $Q$, the following equations of the two-dimensional lock-in technique are evaluated,

$$A_Q(r) = \int dR\, T'(R) e^{iQ\cdot R} e^{-\frac{(r-R)^2}{2\sigma^2}}, \quad (M4)$$

$$|A_Q(r)| = \sqrt{\left(ReA_Q(r)\right)^2 + \left(ImA_Q(r)\right)^2}, \quad (M5)$$

$$\Phi_Q(r) = tan^{-1}\frac{ImA_Q(r)}{ReA_Q(r)}. \quad (M6)$$

where $\sigma$ is a coarse-graining length and $\Phi_Q(r)$ is a spatial phase shift of the modulation at $Q$. In this study, we use $\sigma$ = 5.4 Å. Since $u(r)$ is related to the phase shift $\Phi_Q(r)$ as $Q \cdot u(r) = \Phi_Q(r)$ for $Q=Q_1$ and $Q_2$, then $u(r)$ is obtained by solving equations

$$u(r) = \begin{pmatrix} Q_1 \\ Q_2 \end{pmatrix}^{-1} \begin{pmatrix} \Phi_{Q_1}(r) \\ \Phi_{Q_2}(r) \end{pmatrix}. \quad (M7)$$

Finally, a drift-corrected topograph $T(r)$ is obtained by

$$T(r) = T'(r - u(r)). \quad (M8)$$

By using the same $u(r)$, simultaneously taken electronic structure images $g'(r, \omega)$ are corrected in the same way as (M8)

$$g(r, \omega) = g'(r - u(r), \omega). \quad (M9)$$

$g(r, \omega)$ is the drift-corrected electronic structure map.

The Lawler-Fujita algorithm is key for visualizing intra-unit-cell electronic structure. To obtain this goal, the dataset with atomic resolution and the dataset with high voltage (-1.6 V to 2 V) must be registered with atomic accuracy. The two datasets are measured in the same field of view. The topographs of the two datasets, $T_1'(r)$ in the atomic resolution dataset and $T_2'(r)$ in the high voltage dataset, are used to register the identical atoms in one dataset with those in the other. The Lawler-Fujita algorithm is applied to both datasets to remove the lattice distortions of the image introduced by the distortions of the piezo-tube, and to produce the corrected images of $T_1(r)$ and $T_2(r)$. Subsequently $T_1(r)$ and $T_2(r)$ are registered by spatial translations, whose accuracy is evaluated by the cross-correlation between $T_1(r)$ and $T_2(r)$. The spatial translations correct any small difference in field-of-view between the measurements. $T_1(r)$ and $T_2(r)$ are now registered. All transformation parameters applied to $T_2'(r)$ are subsequently applied to the high-voltage differential conductance map $g'(r, V)$ that is measured simultaneously with the topography. The differential conductance map is $g(r, V)$ registered.

Extended Data Fig. 5 shows two topographs of a $Bi_2Sr_2CaCu_2O_{8+x}$ surface taken with atomic resolution and high voltage. They are corrected using the Lawler-Fujita algorithm and registered with atomic resolution. The precision of the image registration is shown in

Extended Data Fig. 5e. The maximum of the cross-correlation between $T_1(\mathbf{r})$ and $T_2(\mathbf{r})$ coincides with the (0,0) cross-correlation vector. The offset of the two registered images are within three pixels. Taking an uncertainty of half the size of the three pixels provides the precision of the registration method is better than 80 pm everywhere in the whole field-of-view.

To demonstrate that evidence of rotational symmetry breaking is present in the unprocessed charge transfer energy variation $\delta\mathcal{E}'(\mathbf{r})$, Extended Data Fig. 6 compares the unprocessed data of $\delta\mathcal{E}'(\mathbf{r})$ and the processed $\delta\mathcal{E}(\mathbf{r})$. $\delta\mathcal{E}'(\mathbf{r})$ is calculated from the unprocessed $g'(\mathbf{r}, V)$ dataset and $\delta\mathcal{E}(\mathbf{r})$ is calculated from the LF-corrected and registered $g(\mathbf{r}, V)$ dataset. The two images show many features in common. The Fourier transform $\delta\mathcal{E}'(\mathbf{q})$ of the unprocessed data shows anisotropy at the Bragg peaks. The processed data $\delta\mathcal{E}(\mathbf{q})$ shows the same anisotropic Bragg peaks and the background noise is much lower in the drift-corrected data than in the unprocessed data. Thus, the LF algorithm removes the pico-meter distortion due to the STM tip scanner and increases the signal-to-noise ratio. The LF algorithm does not alter the conclusion that $C_4$ symmetry is broken in $\delta\mathcal{E}$.

**E. Intra-unit-cell Charge Transfer Energy Imaging: Repeatability and Efficiency**

Here we first compare the result of multiple experiments to evaluate the repeatability of the phenomena reported. Firstly, the Fourier analysis was repeated on three different field-of-views of the same sample. The results of this analysis are shown in Extended Data Figs. 7-12. The topograph of each FOV is shown in Extended Data Figs. 7, 9, 11 a, the supermodulation can be clearly identified as expected for $Bi_2Sr_2CaCu_2O_{8+x}$ with $p \approx 0.17$. The $T(\mathbf{r})$ is an atomically-resolved topograph that is registered with the differential conductance map $g(\mathbf{r}, V)$. Extended Data Figs. 7, 9, 11 b then show the variations of the charge transfer energy, $\delta\mathcal{E}(\mathbf{r})$. All FOVs show high levels of disorder in these images, and also all span the same energy ranges of -300 meV to +200 meV. The power spectral density (PSD) Fourier transform of the topograph is then shown in Extended Data Figs. 7, 9, 11 c. Highlighted in the blue dashed circle is the Bragg peaks $\mathbf{Q}_y$ and in the red dashed circle is $\mathbf{Q}_x$. The Bragg peaks are of similar magnitude in the topograph, therefore $C_4$ symmetry is preserved indicating that neither the lattice nor the scanning tip produces IUC symmetry breaking. Long-range disorder is seen at $\mathbf{Q} = 0$.

Next, $\delta\mathcal{E}(\mathbf{q})$, the PSD Fourier transform of charge transfer energy variation, is shown in Extended Data Figs. 7, 9, 11 d, again with the Bragg peaks highlighted with dashed circles. In all FOVs it is clear the Bragg peaks at $(0,1)2\pi/a_0$ are more intense than the peaks at $(1,0)2\pi/a_0$, meaning IUC $C_4$ rotational symmetry is broken. Linecuts of $T(\mathbf{q})$ are then taken in Extended Data Figs. 7, 9, 11 e. The red linecut is taken from $\mathbf{q} = (0,0)$ to $(1,0)2\pi/a_0$ and the blue linecut is taken from $\mathbf{q} = (0,0)$ to $(0,1)2\pi/a_0$. These linecuts allow quantitative analysis of Bragg peaks. In all FOVs it is clear $\mathbf{Q}_x \approx \mathbf{Q}_y$ in $T(\mathbf{q})$. Confirming that the tip and lattice preserve C$_4$ symmetry. Extended Data Figs. 7, 9, 11 f are linecuts of $\delta\mathcal{E}(\boldsymbol{q})$, again taken from $\boldsymbol{q} = (0,0)$ to $(1,0)2\pi/a_0$ and the blue linecut is taken from $\boldsymbol{q} = (0,0)$ to $(0,1)2\pi/a_0$. In all FOVs, there is a clear anisotropy between the intensities of the Bragg peaks. Therefore, IUC rotational symmetry is broken in $\delta\mathcal{E}$. The inset is a histogram of the distribution of $\delta\mathcal{E}(\boldsymbol{r})$. The histograms are asymmetric and the peak of the histogram is shifted from 0 in all FOVs.

To sum up Extended Data Figs. 7, 9, 11: the anisotropy in the Bragg peaks of the charge transfer energy variations $\delta\mathcal{E}$ are repeatable in multiple experiments at the same hole density. The $\delta\mathcal{E}$ from independent FOVs show similar statistics. The nematicity is not generated by the crystallography nor the scanning tip. Therefore, the Fourier analysis of $\delta\mathcal{E}$ appears robust and reliable.

Secondly, the IUC charge transfer orbital-order parameter and the oxygen site-specific imaging of the charge transfer are repeated for the three field-of-views in Extended Data Figs. 8, 10, 12, respectively. Extended Data Figs. 8, 10, 12 a report the Ising domains for the data in Extended Data Figs. 7, 9, 11, respectively. $D_\mathcal{E}(\boldsymbol{r})$ is the intra-unit-cell d-symmetry breaking of the charge transfer energy variations and its calculation details of $D_\mathcal{E}(\boldsymbol{r})$ are discussed in the following Section F. The dashed circle shows the $r$-space radius of the gaussian smoothing used in calculating $D_\mathcal{E}(\boldsymbol{r})$. The three FOVs show similar disordered Ising domains. The relative strength between the two domains is defined as the area ratio $A_{red}/A_{blue}$ which is approximately $2.1 \pm 0.2$. This is consistent with the relative strength between the two Bragg peaks in the power spectral density Fourier transform of $\delta\mathcal{E}$, $\frac{\delta\mathcal{E}(\boldsymbol{Q}_y)}{\delta\mathcal{E}(\boldsymbol{Q}_x)} \sim 1.9 \pm 0.3$. This result demonstrates that the relative preponderance of the two orbital order domains is approximately 2:1 in the 17% hole-doped sample. The observed preponderance of the alpha domain relative to the beta domain in our study could be attributed to several factors that influence the distribution of electronic phases in cuprates. However, one simple explanation presents itself. If orbital ordering is a stable thermodynamics phase lowering the free energy density, one might expect it to be homogenous throughout a perfect crystal. However, when the locations of oxygen dopant ions are imaged simultaneously with $N_\mathcal{E}(\boldsymbol{r})$ (Methods Section H), they occur with enhanced probability proximate to the orbital ordering domain walls. This implies that the quenched disorder from dopant ions pins the orbital ordered domains and, if so, some domains of the opposite sign to the predominant order will be stabilized. In that case, one naturally expects that one domain (alpha) is more abundant than the other (beta).

Extended Data Figs. 8, 10, 12 f display the oxygen site-specific order parameter $N_\mathcal{E}(\boldsymbol{r})$ images (Equation 3 in the Main Text). The spatial distribution of $N_\mathcal{E}(\boldsymbol{r})$ is highly similar to that of $D_\mathcal{E}(\boldsymbol{r})$ with a cross-correlation coefficient of 0.93. This focuses attention on the charge transfer symmetry breaking mainly occurring on the planar oxygen sites.

The microscopic structure inside the Ising domains is visualized from the unit-cell-averaged images of each domain, as shown in Extended Data Fig. 8, 10, 12 b-c. The unit cell average from the $N_{\mathcal{E}}(r) > 5$ meV region is indicated in red, and the unit cell average from the $N_{\mathcal{E}}(r) < -5$ meV is indicated in blue. Clearly the charge transfer energy variations break the rotational symmetry inside the unit cells. Because the number of unit cells in $N_{\mathcal{E}}(r) > 5$ meV domain is almost twice the other domain, the signal-to-noise ratio is higher in the unit cell averaging than results in $\delta\mathcal{E}_B^R(r)$. The unit-cell-averaged images of the topograph $T(r)$ are then presented in Extended Data Figs. 8, 10, 12 d-e as a reference of Extended Data Figs. 8, 10, 12 b-c. The tip preserves IUC rotational symmetry and the crystallography of the $CuO_2$ unit cell does not break rotational symmetry.

The histogram of $N_{\mathcal{E}}(r)$ in Extended Data Figs. 8, 10, 12 g are asymmetric and the peak of the histogram is shifted from $N_{\mathcal{E}} = 0$. This observation is consistent with Extended Data Figs. 8, 10, 12 f that the preponderance of the $N_{\mathcal{E}}(r) > 0$ domain is stronger than the $N_{\mathcal{E}}(r) < 0$ domain. The RMS of the IUC energy splitting between the $O_x$ and $O_y$ sites ranges from 20 meV to 30 meV.

Third, we demonstrate directly that the rotational symmetry breaking we report is a true property of the charge transfer gap from unprocessed data (Extended Data Figs. 13-15). Three representative examples of this from three different but perfectly typical FOVs as now presented in Extended Data Figs. 13-15. The randomly chosen locations of the three FOVs are presented in Extended Data Fig. 16. These same phenomena are observed omnipresent throughout the total FOV of the experiment, meaning that orbital ordering is universal.

We show the topography, the orbital-order order parameter $N_{\mathcal{E}}(r)$, and 20 unprocessed $dI/dV$ point spectra and from each FOV. Extended Data Figs. 13-15a are the topographs showing the FOVs where the charge transfer measurements are taken. The locations of five $CuO_2$ unit cells consisting of 10 different oxygen sites are highlighted (blue and red dots). Figs. 13-15b shows the nematic domains $N_{\mathcal{E}}(r)$ in this FOV. Figs. 13-15c shows Domain 1 (red) where Ox spectra are shifted by -50 meV ~ -30 meV with respect to the Oy spectra inside the same unit cell. The charge transfer gap on the Ox sites (red) are higher than the gap on the Oy site (blue) in this domain. Domains 2 (blue) show Oy spectra are shifted by -50 meV ~ -30 meV with respect to the Ox spectra (Figs. 13-15d), which means that the Oy sites have higher charge transfer energy than the Ox sites in this domain. The two domains are separated by a domain wall. Such domains are observed throughout unprocessed data showing the intra-unit-cell symmetry breaking and the Ising domains of orbital ordering. While presenting such unprocessed spectra comparing $O_x$ and $O_y$ from each unit cell clearly shows the presence of orbital ordering and Ising domains, it would be cumbersome, both for analysis and presentation, to do so for the 255 unit cells contained in Main Text Fig. 2. Furthermore, the locations and variations of the Ising domains would become obscured. By presenting $N_\mathcal{E}(r)$ instead, we can effectively visualize and quantify variations of intra-unit-cell energy splitting. These $N_\mathcal{E}(r)$ maps are particularly suited to our study as they vividly demonstrate the existence of Ising domains and how they are influenced by the presence of interstitial oxygen dopant atoms, Main Text Fig. 3e.

The repeatability of charge transfer energy splitting in three additional experiments using a different setup condition (-750 mV, 25 pA) in the unprocessed data demonstrates that the evidence of orbital ordering is independent of setup conditions and data analysis, and orbital ordering is a robust experimental result. The two datasets are complementary, but it's crucial to emphasize the indispensable role of the original dataset in measuring charge transfer energy variations. Measured at 85 GΩ and spanning a wide energy range of -1.6 V to 2 V, this dataset provides a precise and rigorous mapping of charge transfer energy variations. This is a direct indicator of orbital ordering, a critical aspect that underpins the entire studyThis independent experiment robustly and repeatedly shows the -50 meV ~ -30 meV energy splitting between $O_x$ and $O_y$ sites within the unit cell in the -750 mV and 25 pA setpoint experiments, all in unprocessed data. Thus, in $Bi_2Sr_2CaCu_2O_{8+x}$ intra-unit cell rotational symmetry breaking at the charge transfer energy scale definitely does occur on the intra-unit-cell oxygen sites, and this effect occurs oppositely in two Ising domains.

**F. Visualizing Intra-unit-cell Charge Transfer Order Parameter**

This section introduces the techniques to visualize the intra-unit-cell charge transfer energy variations. The FOV in the main text contains ~500 each of individually resolved Cu sites. Extended Data Fig. 17a shows the topograph with Cu sites indicated by red circles. The Cu site $\boldsymbol{R}_{ij}$ is identified using the Lawler-Fujita phase definition algorithm. Extended Data Fig. 17b is the simultaneous charge transfer image with the locations of the Cu sites overlaid. Each $CuO_2$ unit cell is defined by a central Cu site and four Cu atoms at the corner (Inset of Extended Data Fig. 19a). The $CuO_2$ unit cell is segregated into four quadrants $x+$, $x-$, $y+$ and $y-$. From the four lists of lattice positions, four masking images are created, $M_{x+}(r)$, $M_{x-}(r)$, $M_{y+}(r)$, and $M_{y-}(r)$. These are discrete images consisting of triangular masks with the origin at the Cu site $\boldsymbol{R}_{ij}$ and two edges along the diagonal of the $CuO_2$ unit cell. The pixel inside the mask is assigned to value 1 and the rest pixels are the value 0. Typically, 10 pixels are used per quadrant. The four-quadrant segregated images of the charge transfer energy $\delta\mathcal{E}(r)$ are then given by

$$\delta\varepsilon_{x+}(r) = \delta\varepsilon(r)M_{x+}(r) \quad \text{(M10)}$$
$$\delta\varepsilon_{x-}(r) = \delta\varepsilon(r)M_{x-}(r) \quad \text{(M11)}$$
$$\delta\varepsilon_{y+}(r) = \delta\varepsilon(r)M_{y+}(r) \quad \text{(M12)}$$
$$\delta\varepsilon_{y-}(r) = \delta\varepsilon(r)M_{y-}(r) \quad \text{(M13)}$$

Extended Data Fig. 17c-f show the quadrant-segregated images of $Bi_2Sr_2CaCu_2O_{8+x}$. The values from the four quadrants are registered by symmetrizing with respect to either the central Cu site or the diagonal of the CuO$_2$ unit cell. Subsequently the d-symmetry breaking order parameter $D_\varepsilon(\boldsymbol{R}_{ij})$ is calculated for each unit cell $\boldsymbol{R}_{ij}$ using equation M14. This is a discrete image sampled at one quadrant inside each unit cell. Gaussian blur is applied to this $D_\varepsilon(\boldsymbol{R}_{ij})$ to generate a continuous spatially averaged image where the nematic Ising domains are visualized.

$$D_\varepsilon(\boldsymbol{R}_{i,j}) = (\delta\varepsilon_{x+} + \delta\varepsilon_{x-})_{i,j}/2 - (\delta\varepsilon_{y+} + \delta\varepsilon_{y-})_{i,j}/2 \qquad \text{(M14)}$$

To determine the fraction of the total area occupied by the two-level systems (TLS) in the sample, and to validate the estimated domain wall cut-off at $|N_\varepsilon| = 5$ meV, we calculate the domain boundary trajectory and broaden it by the correlation length. The correlation length of $N_\varepsilon(\boldsymbol{r})$ is used to estimate the distance from the domain walls where these TLS are likely to be thermally active. First, the autocorrelation of $N_\varepsilon(\boldsymbol{r})$ is given by

$$A(\boldsymbol{R}) = \int_{-\infty}^{\infty}[N_\varepsilon(\boldsymbol{r}) - \overline{N}_\varepsilon] \times [N_\varepsilon(\boldsymbol{r} + \boldsymbol{R}) - \overline{N}_\varepsilon]d\boldsymbol{r} \qquad \text{(M15)}$$

where $\boldsymbol{R}$ is the location being integrated at, while $\boldsymbol{r}$ is varied. This is shown in Extended Data Fig. 18a. The decay length of this autocorrelation function is defined as the correlation length $\xi$. This can be determined by plotting the autocorrelation as a function of radial angle and then fitting, where $C$ is a fitting constant

$$A_\theta(r) = Ce^{-\xi r} \qquad \text{(M16)}$$

A plot of the autocorrelation as a function of radius is shown in Extended Data Fig. 18b and the inset shows the corresponding correlation length $\xi$ as a function of radial angle. Averaging $\xi$ over radial angle results in an averaged correlation length of $\bar{\xi} = 3.1$ Å. The domain boundaries are then determined using $\frac{\max(|N_\varepsilon(\boldsymbol{r})|)}{|N_\varepsilon(\boldsymbol{r})|} \gg 1$, these boundaries are shown in Extended Data Fig. 18c. Extending the domain boundaries by the correlation length of 3.1 Å gives rise to the area of the domain walls. The domain wall regions host plausibly thermally active TLS quadrupoles (Extended Data Fig. 18d), and the area fraction of approximately 25% TLS is found inside these domain walls.

In the main text the domain walls are defined as $|N_\varepsilon(\boldsymbol{r})| < 5$ meV (the white area defined in Fig. 3d). This leads to an area fraction of 25%, consistent with the method used above. Thus, the cut-off of $|N_\varepsilon(\boldsymbol{r})| < 5$ meV is mathematically validated using the above algorithm. Further experimental work focusing on the dynamics of domain walls, specifically observing how they change over time and respond to varying temperatures, would significantly deepen our understanding of the orbital ordering domain walls and the thermally activated two-level systems within them. One method to do so would be to use the STM tip to detect flicker noise in the tunnelling current. The flicker noise of the two-level systems could be extracted using a cryogenic amplifier as described for example in Refs. 35,36. Their fluctuations could be increased as a function of temperature. This is an intriguing experimental prospect.

The unit cells inside one Ising domain are averaged to increase the signal-to-noise ratio of the microstructure of the IUC symmetry breaking order[37]. Firstly, each $CuO_2$ unit cell is identified in the topography $T(r)$ (gray grid in Extended Data Fig. 19a). Each unit cell has a size of 10×10 pixels and is labelled as $I(x, y, n)$. The image series $I(x, y, n)$ is categorized into two datasets $I_1(x, y, n)$ and $I_2(x, y, n)$. Where $I_1$ is inside a positive domain with $N(r) > 5\ meV$ and $I_2$ is inside a negative domain with $N_\varepsilon(r) < -5\ meV$ (Extended Data Fig. 19b). The images $I_{1,2}(x, y, n)$ have lateral dimensions $x$ and $y$ and a counting index $n$. $T(r)$ has already been atomically registered and the lattice displacements have been corrected using the Lawler-Fujita algorithm, making each lattice square and periodic. The image set of $I_{1,2}(x, y, n)$ ) is made up with $n$ images of $I_{1,2}(x, y)$. The images are cropped independently from $T(r)$, where each image has a size of 11 × 11 pixels (5.9 × 5.9 Å) and corresponds to one $CuO_2$ unit cell rotated by 45°. The single $CuO_2$ unit cell is defined by the location of Bi/Cu atoms (Extended Data Fig. 19).

**G. Oxygen-site Specific Charge Transfer Order: Real Space and Reciprocal Space**

From our analysis Extended Data Figs. 8, 10, 12 b, c we see that the key difference in the CT energy is between oxygen sites, for this reason we introduce the oxygen-site specific charge transfer nematic order parameter. The dual real and reciprocal-space representation of the sublattices are constructed. The real-space atomic decomposition of Cu, O$x$+, O$x$-, O$y$+ and O$y$- sublattices is given by

$$\boldsymbol{R}_{ij}^{Cu} = i\hat{\boldsymbol{x}} + j\hat{\boldsymbol{y}} \tag{M17}$$

$\boldsymbol{R}_{ij}^{Cu}$ is the coordinate of a Cu site, therefore we extract the value of the charge transfer energy variation, $\delta\varepsilon_{Cu}(r)$, on this coordinate via M18

$$\delta\varepsilon_{Cu}(r) = \delta\varepsilon(\boldsymbol{R}_{ij}^{Cu}) \tag{M18}$$

$\boldsymbol{R}_{ij}^{Cu} + ax/2$ is the coordinate of an Oxygen site in the positive $x$ direction and vice versa for $\boldsymbol{R}_{ij}^{Cu} - ax/2$, therefore we extract the value of the charge transfer energy variation, $\delta\varepsilon_{x\pm}^{R}(r)$, on this coordinate via M19

$$\delta\varepsilon_{x\pm}^{R}(r) = \delta\varepsilon_{Cu}(r \pm a\hat{x}/2) \Rightarrow \delta\varepsilon_{x\pm}^{R}(\boldsymbol{R}_{ij}) = \delta\varepsilon_{Cu}(\boldsymbol{R}_{ij} \pm a\hat{x}/2) \tag{M19}$$

$\boldsymbol{R}_{ij}^{Cu} + ay/2$ is the coordinate of an Oxygen site in the positive $y$ direction and vice versa for $\boldsymbol{R}_{ij}^{Cu} - ay/2$, therefore we extract the value of the charge transfer energy variation, $\delta\varepsilon_{y\pm}^{R}(r)$, on this coordinate via M20

$$\delta\varepsilon^R_{y\pm}(\boldsymbol{r}) = \delta\varepsilon_{Cu}(\boldsymbol{r} \pm a\hat{\boldsymbol{y}}/2) \Longrightarrow \delta\varepsilon^R_{y\pm}(\boldsymbol{R}_{ij}) = \delta\varepsilon_{Cu}(\boldsymbol{R}_{ij} \pm a\hat{\boldsymbol{y}}/2) \quad \text{(M20)}$$

The $\boldsymbol{R}^{Cu}_{ij}$ are the set of direct lattice vectors of the $CuO_2$ lattice with lattice spacing $a$. The oxygen sublattices are displaced by half a unit cell so that the origin of the oxygen sublattice is located on an oxygen site.

The sublattice decomposition using Fourier analysis[38] in the reciprocal-space is given by

$$\delta\varepsilon^Q_{x\pm}(x,y) = \delta\varepsilon(\boldsymbol{r})cos(Q_x * x \pm \pi)cos(Q_y * y) \Longrightarrow \delta\varepsilon^Q_{x\pm}(\boldsymbol{R}_{ij}) = \delta\varepsilon^Q_x(\boldsymbol{R}_{ij} \pm a\hat{\boldsymbol{x}}/2) \quad \text{(M21)}$$
$$\delta\varepsilon^Q_{y\pm}(x,y) = \delta\varepsilon(\boldsymbol{r})cos(Q_x * x)cos(Q_y * y \pm \pi) \Longrightarrow \delta\varepsilon^Q_{y\pm}(\boldsymbol{R}_{ij}) = \delta\varepsilon^Q_y(\boldsymbol{R}_{ij} \pm a\hat{\boldsymbol{y}}/2) \quad \text{(M22)}$$

Here the sublattice is a product of two cosine window functions because the product has a more localized focus on the sublattice than the sum of two cosine window functions. The negative values in the window functions $cos(Q_x * x \pm \pi)cos(Q_y * y)$ and $cos(Q_x * x)cos(Q_y * y \pm \pi)$ are set to 0. Like the real-space construction, the oxygen sublattices are displaced by half a unit cell and one oxygen atom is set as the origin of this sublattice. The sublattice decomposition from the reciprocal space rules out the error of identifying the position of the atomic sites in the real space.

The real space nematic order parameter can now be constructed

$$N^R_\varepsilon(\boldsymbol{R}_{ij}) = \delta\varepsilon^R_x(\boldsymbol{R}_{ij}) - \delta\varepsilon^R_y(\boldsymbol{R}_{ij}) \quad \text{(M23)}$$

$$N^R_\varepsilon(\boldsymbol{R}_{ij}) = \frac{[\delta\varepsilon_{Ox}(\boldsymbol{R}_{ij}+a\hat{\boldsymbol{x}}/2)+\delta\varepsilon_{Ox}(\boldsymbol{R}_{ij}-a\hat{\boldsymbol{x}}/2)]}{2} - \frac{[\delta\varepsilon_{Oy}(\boldsymbol{R}_{ij}+a\hat{\boldsymbol{y}}/2)+\delta\varepsilon_{Oy}(\boldsymbol{R}_{ij}-a\hat{\boldsymbol{y}}/2)]}{2} \quad \text{(M24)}$$

Similarly, the nematic order parameter is calculated from the reciprocal space analysis
$$N^Q_\varepsilon(\boldsymbol{R}_{ij}) = \delta\varepsilon^Q_x(\boldsymbol{R}_{ij}) - \delta\varepsilon^Q_y(\boldsymbol{R}_{ij}) \quad \text{(M25)}$$
The images of the segregated $O_x$ and $O_y$ sublattices $\delta\varepsilon_x(\boldsymbol{R}_{ij})$ and $\delta\varepsilon_y(\boldsymbol{R}_{ij})$, the discrete images of the nematic order parameter $N_\varepsilon(\boldsymbol{R}_{ij})$ and the continuous image $N_\varepsilon(\boldsymbol{r})$ are presented in Extended Data Fig. 20. The sublattice decomposition calculated from real space and from reciprocal space are almost identical, which validates the sublattice decomposition algorithm.

While the order parameters $N_\varepsilon(\boldsymbol{r})$ and $D_\varepsilon(\boldsymbol{r})$ originate from two slightly different physical definitions, they are clearly similar visually, as expected. The cross-correlation coefficient of these two images is 0.93 (where 1 means the two images are identical) this confirms that both methods are able to visualize these orbital ordering Ising domains.

## H. Dopant Oxygen Ion Pinning of Orbital Ordered Domains

To explore causes for the obvious heterogeneity on the orbital order domains, we search for the dopant oxygen ions and their relationship with the domains. Each dopant oxygen ion is identified as a maximum located at -900 meV in the differential conductance spectrum (inset of Extended Data Fig. 21a). A total of ~70 oxygen dopants are found in the differential conductance map $g(r, -900$ meV$)$ with an image size of 17 × 17 nm$^2$ (Extended Data Fig. 21a). The locations of the oxygen dopant are overlaid onto the simultaneously measured IUC oxygen-specific nematic order parameter $N_\mathcal{E}(r)$ in Extended Data Fig. 21d. Visually we see that the oxygen dopants are near the $N_\mathcal{E}$ domain walls (yellow contours).

The distance $d_{\text{dopant}}$ between each oxygen dopant and their nearest location on the domain walls is calculated and a distribution is shown in Extended Data Fig. 21g. To validate that the oxygen dopants are located near the $N_\mathcal{E}$ domain walls, the distribution of $d_{\text{dopant}}$ is compared to the distance $d_{\text{random}}$ between randomly generated points and the domain walls. The random points are generated from a two-dimensional Poisson disc sampling function and the random points are separated from each other. The random points have no spatial correlation with the orbital ordering domains. The expected averaged distance $d_{\text{random}}$ to the domain walls is calculated and compared to $d_{\text{dopant}}$ in Extended Data Fig. 21g.

The distribution of $d_{\text{dopant}}$ is different from $d_{\text{random}}$ from two aspects. While $d_{\text{dopant}}$ distribution has a sharp peak at 1.6 Å, $d_{\text{random}}$ distribution has a blunt plateau. The deviation in the distance distribution indicates clearly that the oxygen dopants are located near the domain walls of the orbital order domains, providing statistical evidence that they are pinning the nematic domains.

We repeated the above statistical analysis for three independent FOVs (Extended Data Fig. 21a-c). Extended Data Fig. 21f is chosen to be presented in the main text Fig. 3e. The sum of the three histograms (Extended Data Fig. 21g-i) are presented in the main text Fig. 3f. A total of 237 oxygen dopants are studied in the total histogram.

The field of view used in the main text contains 70 dopant atoms. The dopant radius is quantified as the spatial decay length of the dopant resonance[1] away from the dopant site. We take spatially averaged linecuts over typical dopant maxima and fit with an exponential decay function. The average decay length gives us a dopant size of approximately 2 Å. A ratio of dopant area to the total field-of-view area can be calculated to be on the order of 10%. Therefore, dopant atoms do not make a significant difference to overall $\delta\mathcal{E}(r)$ imaging procedures.

**I. Sample Quality Characterization**

It is important to consider sample quality, to ensure a high-quality measurement was taken. The single crystal of $Bi_2Sr_2CaCu_2O_{8+x}$ sample studied throughout this study is shown in Extended Data Fig. 22a. The sample is 17% hole-doped. This is an isotope substituted $Bi_2Sr_2CaCu_2O_{8+x}$ such that the oxygens ($^{16}O$) in the sample are almost fully substituted by $^{18}O$. The sample quality is characterized by a magnetic susceptibility measurement *M/H* versus temperature across the superconducting transition, shown in Extended Data Fig. 22b. *M* is the magnetization and *H* is magnetic field strength. Here, the samples magnetic susceptibility was measured from 50 to 110 K, with a sharp transition seen at 88 K, whose behavior indicates a superconducting transition. Therefore, from this data it is clear a high-quality sample of $Bi_2Sr_2CaCu_2O_{8+x}$ was studied.

**J. Orbital ordering domains are unaffected by supermodulation.**

To further epitomize the lack of effect the supermodulation has on ***Q*** = 0 orbital ordering, we have compared the topographs and the orbital ordering domains obtained with and without the supermodulation from three independent FOVs (Extended Data Fig. 23). The three FOVs are randomly chosen from Extended Data Fig. 21. The first column (Extended Data Fig. 23a, e, i) shows the topographs where the supermodulation is removed via Fourier filtering at the ***Q***$_{SM}$ peak. Their orbital ordering domains are presented in the third column (Extended Data Fig. 23c, g, k). The second column presents the topographs where the supermodulation is kept in the analysis (Extended Data Fig. 23b, f, j). Although the atomic structure is strongly modulated by the supermodulation, their orbital ordering domains in the fourth column (Extended Data Fig. 23d, h, l) remain virtually the same as the orbital ordering domains where the ***Q***$_{SM}$ peak is filtered (the third column of Extended Data Fig. 23).

Our experimental results align well with our expectations as illustrated in Main Text Fig. 2a, where the bulk incommensurate crystal supermodulation is clearly visible. This supermodulation is oriented at 45°, serving as the mirror plane between the x and y axes, consistent with previous findings. Importantly, this supermodulation does not impact the intra-unit-cell symmetry breaking of $\delta\mathcal{E}(r)$ discussed in this paper, owing to the specific symmetry involved[34]. In conclusion the orbital ordering domains are independent of the supermodulation and thus the orbital ordering is not a vestige of the crystal supermodulation but a true property of $Bi_2Sr_2CaCu_2O_{8+x}$ electronic structure.

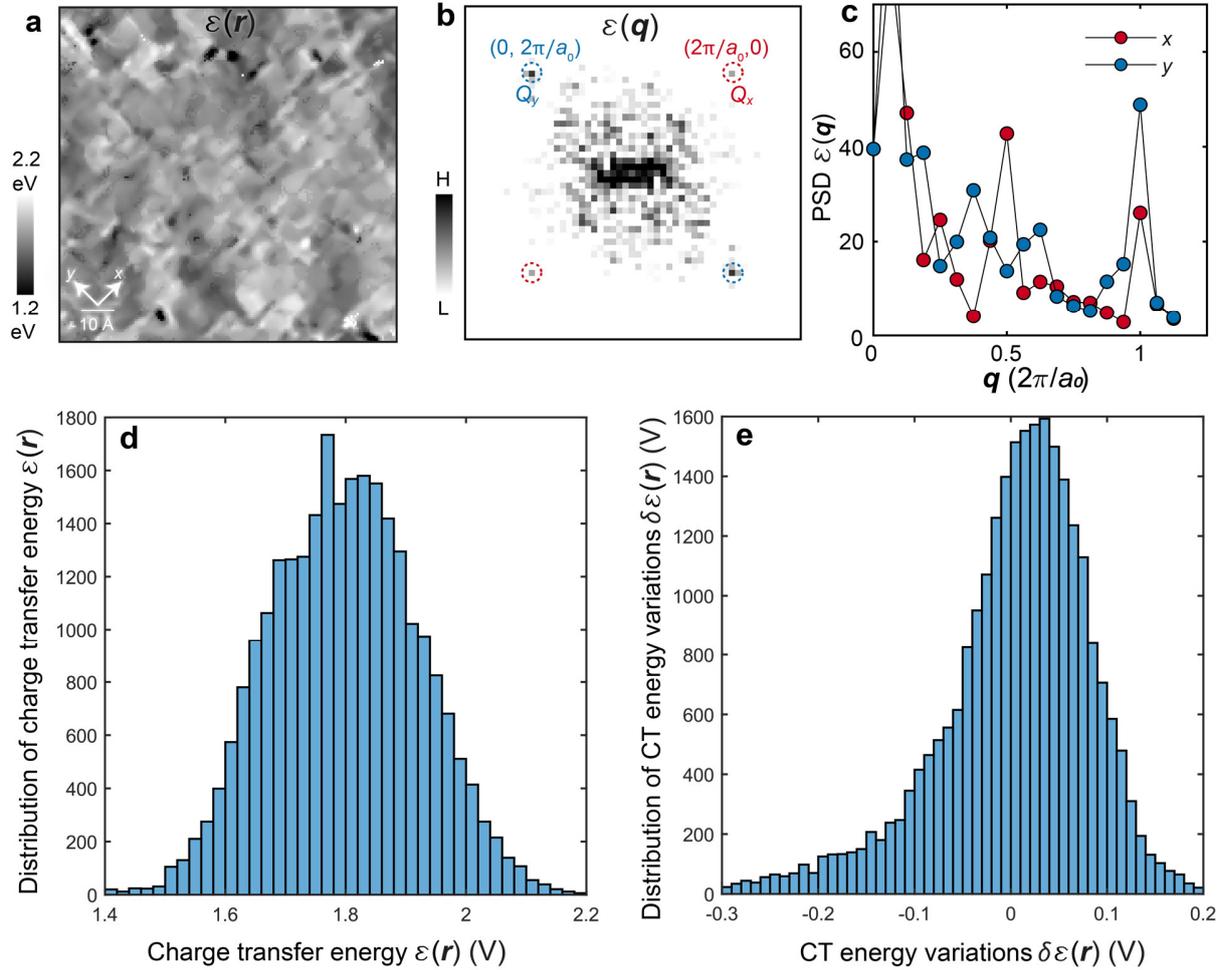

**Extended Data Fig. 1. Comparison of the charge transfer energy $\mathcal{E}(r)$ and the charge transfer energy variations $\delta\mathcal{E}(r)$.** (**a**) Image of the charge transfer energy $\mathcal{E}(r)$ calculated from Eqn. M1. The supermodulation (~150 meV) wavevector $Q_{SM}$ is masked in (a) such that the intra-unit-cell features (~50 meV) are revealed. (**b**) Fourier transform of $\mathcal{E}(r)$. (**c**). Linecut from (0,0) to the Bragg peaks in $\mathcal{E}(r)$, where the ratio of the Bragg peaks intensity $\frac{Q_y}{Q_x} = 1.9$ indicates IUC symmetry breaking in charge transfer energy $\mathcal{E}(r)$. (**d**) Statistical distribution of $\mathcal{E}(r)$. (**e**) Statistical distribution of $\delta\mathcal{E}(r)$. The narrower distribution in $\delta\mathcal{E}(r)$ demonstrates an improvement in signal-to-noise ratio over the $\mathcal{E}(r)$ algorithm. From ED. Fig. 1, the histogram of $\delta\mathcal{E}(r)$ has a narrower σ than the histogram of $\mathcal{E}(r)$, demonstrating the signal-to-noise ratio is higher in the algorithm used here to measure the variations in charge transfer energy $\delta\mathcal{E}(r)$. Thus this $\delta\mathcal{E}(r)$ algorithm with higher SNR was used throughout this paper.

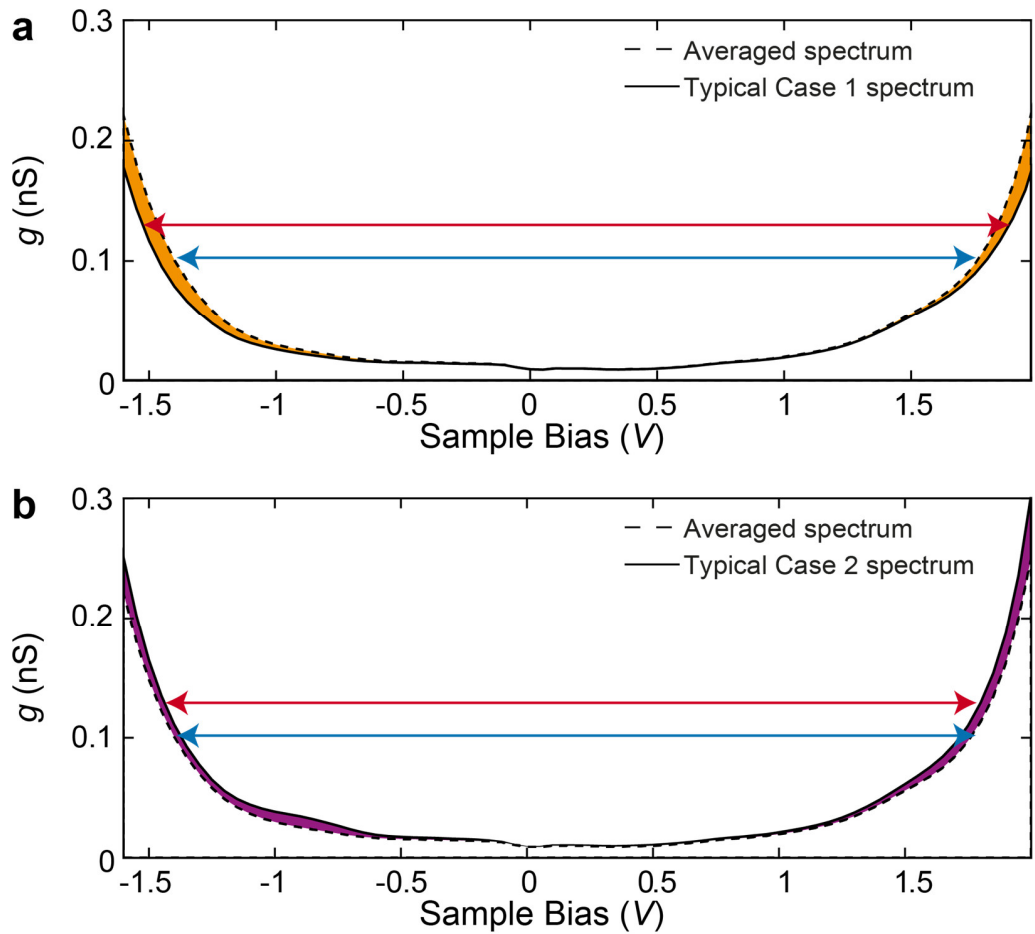

**Extended Data Fig. 2. Typical examples of the splitting of the charge transfer energy.** (**a**) shows a case when the energy variation is positive, $\delta\mathcal{E} > 0$. (**b**) shows a case when the energy variation is negative, $\delta\mathcal{E} < 0$.

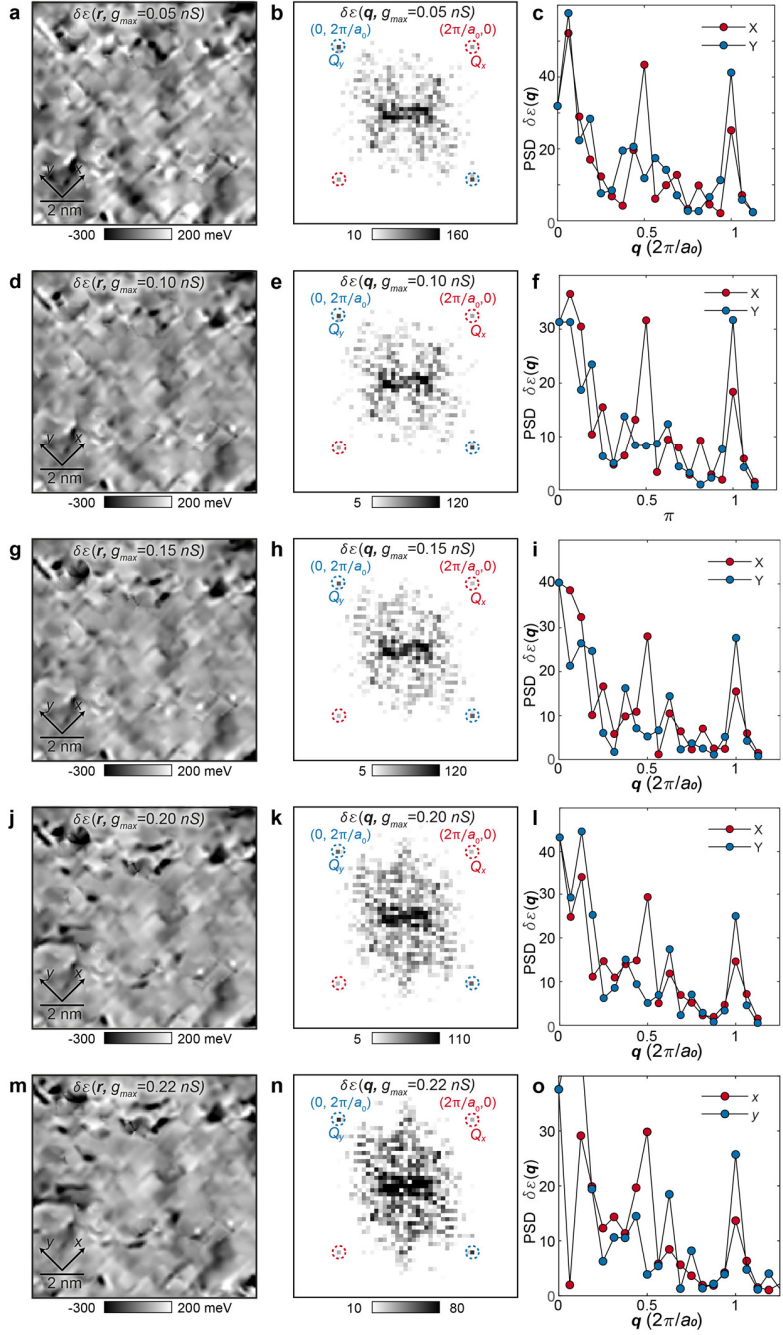

**Extended Data Fig. 3. Calculation of $\delta\mathcal{E}(\mathbf{r})$ at various values of $g_{max}$.** The first column shows $\delta\mathcal{E}(\mathbf{r})$ with $g_{max}$ = 0.05 nS, 0.10 nS, 0.15 nS, 0.20 nS, 0.22 nS, where $g_{max}$ = 0.22 nS is the integration of the full range of spectrum. Strong spatial correlation is seen across all $\delta\mathcal{E}(\mathbf{r})$ images. The middle column is $\delta\mathcal{E}(\mathbf{q})$, the corresponding PSD Fourier transform to the $\delta\mathcal{E}(\mathbf{r})$. The Bragg peaks are highlighted in the dashed circle and all show anisotropy between the intensity of the Bragg peaks $\mathbf{Q}_x$ and $\mathbf{Q}_y$. The $\mathbf{Q} = 0$ disorder increases as $g_{max}$ increases. The third column shows linecuts of $\delta\mathcal{E}(\mathbf{q})$ in the $\mathbf{Q}_x$ and $\mathbf{Q}_y$ directions, showing that the Bragg peaks are always anisotropic at whatever chosen integration conductance $g_{max}$.

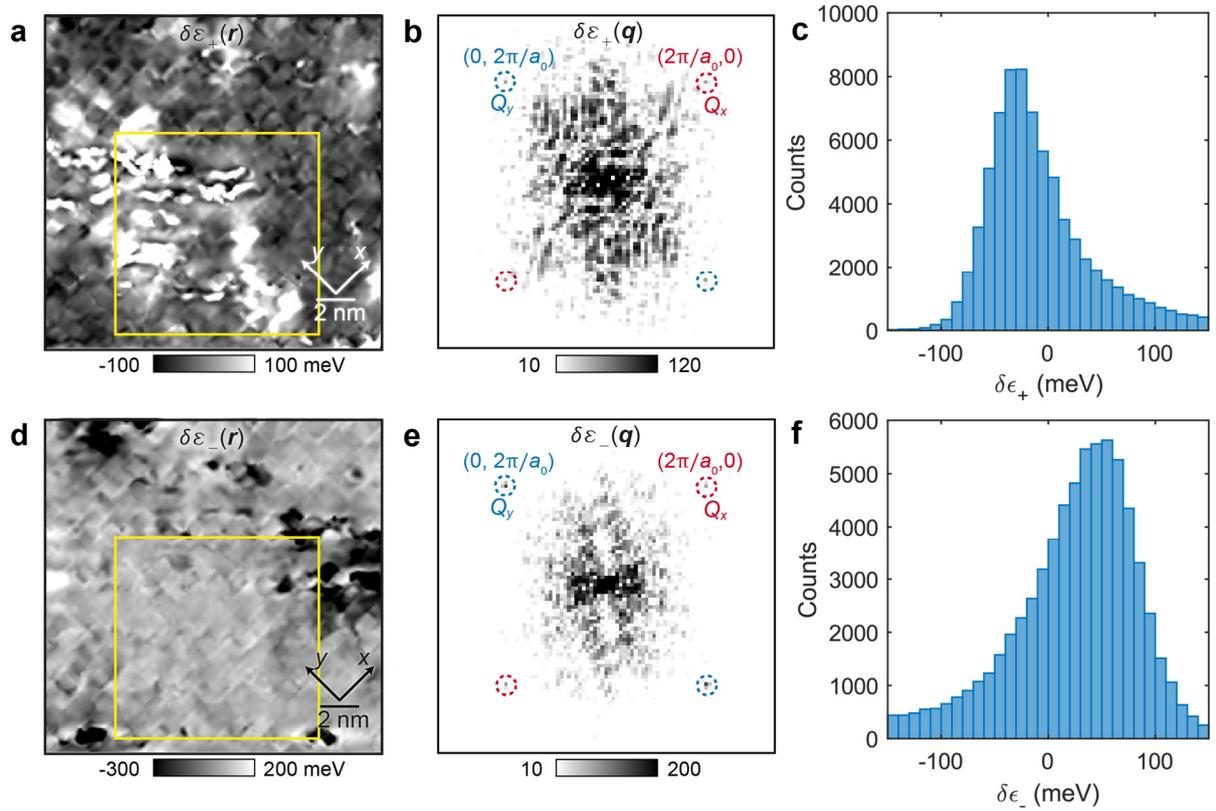

**Extended Data Fig. 4. Charge transfer energy variations of the empty states and filled states.** (**a**) The charge transfer variation of the empty states $\delta\mathcal{E}_+(r)$ and (**b**) its corresponding Fourier transform $\delta\mathcal{E}_+(q)$ showing $\delta\mathcal{E}_+(r)$ is dominated by disorder. (**c**) Histogram showing the distribution of charge transfer variation for the empty states. (**d**) The charge transfer variation of the filled states $\delta\mathcal{E}_-(r)$ and (**e**) its corresponding Fourier transform $\delta\mathcal{E}_-(q)$, showing the filled states are nematic. (**f**) is a histogram showing the distribution of the charge transfer variation for the filled states, clearly this is shifted opposite to the empty states. The yellow box in (a) and (d), is the more ordered region presented in the main text.

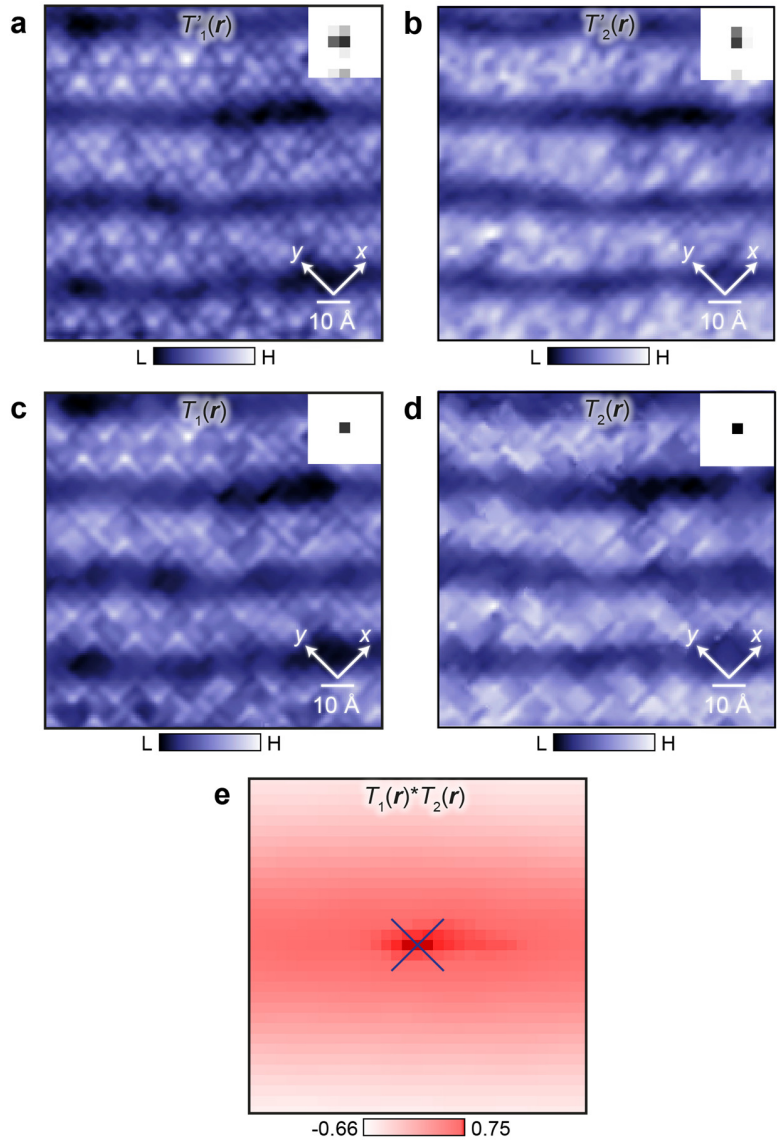

**Extended Data Fig. 5. Spatially registered topographs of $Bi_2Sr_2CaCu_2O_{8+x}$ taken at (a & c) atomic resolution and (b & d) high voltage.** (**a** & **b**) Show two unprocessed low and high voltage topographs in the same field of view. The low-voltage topograph (a) has atomic resolution and it allows Bi atoms to be identified. The high voltage topograph in (b) was taken simultaneously with the electronic structure $g(r,V)$. The insets show the broad Bragg peaks of its topo. (**c** & **d**) show the same topographs after Lawler-Fujita correction. Insets show sharp Bragg peaks indicating that piezo drift has been corrected. (a & c) were acquired with a 30 GΩ junction resistance, at -750 mV tip-sample bias. (b &d) were acquired with an 85 GΩ junction resistance, at 600 mV tip-sample bias. (**e**) Zoom-in image (center) of the cross-correlation between $T_1(r)$ and $T_2(r)$. The maximum (annotated by the cross) has a width of 3 pixels. The registration precision is better than 80 pm (equivalent to 1.5 pixels).

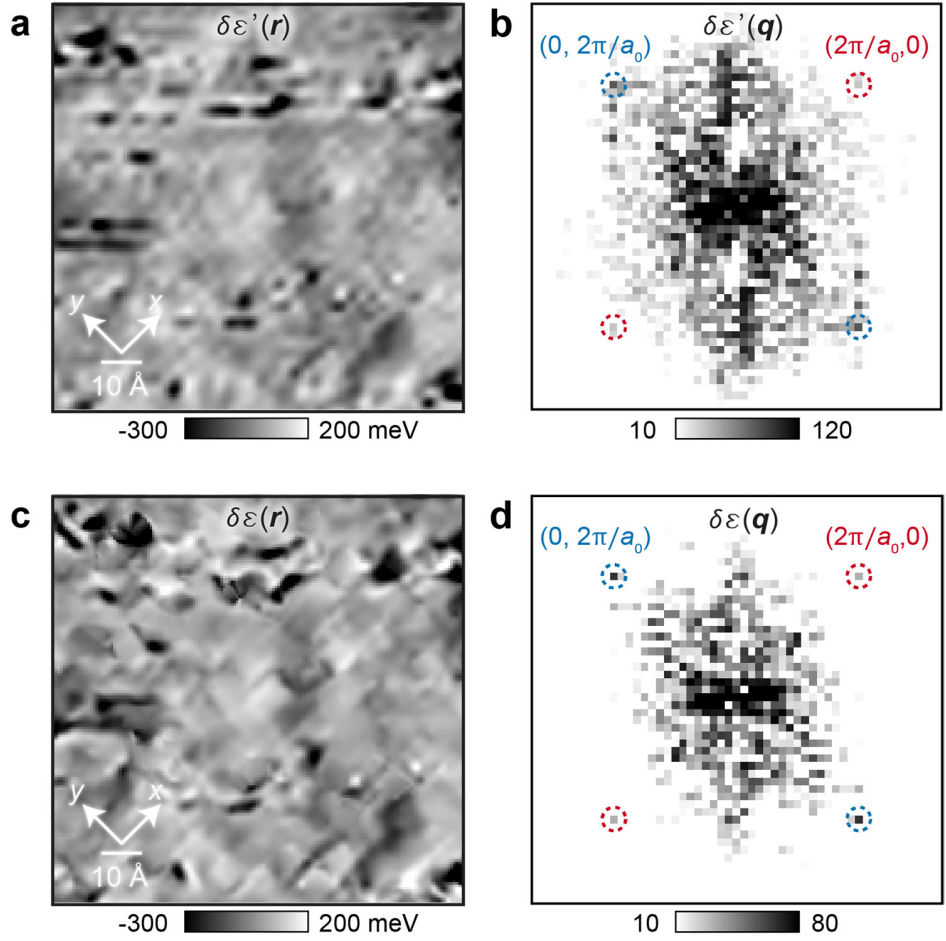

**Extended Data Fig. 6. The variation of the charge transfer energy before and after Lawler-Fujita correction.** (**a**) is an unprocessed map of $\delta\mathcal{E}'(r)$ and (**b**) is the corresponding PSD Fourier transform. $\delta\mathcal{E}'(q)$ shows anisotropy between its Bragg peaks, which are marked by dashed circles, with $\frac{Q_y}{Q_x} = 2.0$. (**c**) is $\delta\mathcal{E}(r)$ after correcting for piezo-drift and (**d**) the corresponding PSD Fourier transform. The Bragg peaks are single pixel and IUC symmetry breaking remains present at the Bragg peaks with $\frac{Q_y}{Q_x} = 1.9$. The disorder at $q = 0$ has been noticeably reduced.

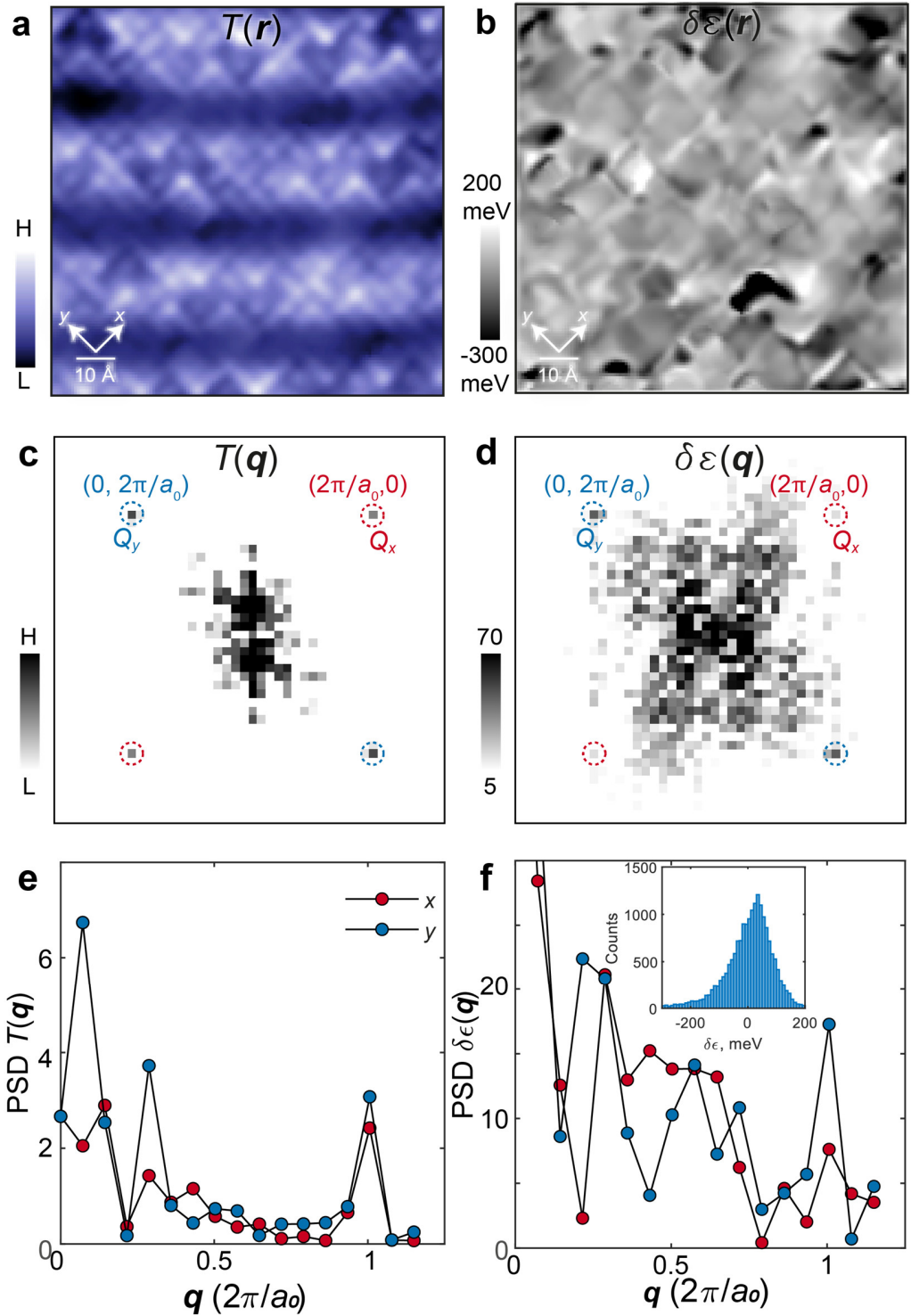

**Extended Data Fig. 7. Fourier analysis of charge-transfer energy variations map $\delta\mathcal{E}(r)$ in another FOV.** (e) $T(q)$ shows the ratio of the transverse averaged intensity at the Bragg peaks is $\frac{Q_y}{Q_x} = 1.3$. (f) $\delta\mathcal{E}(q)$ shows $\frac{Q_y}{Q_x} = 2.3$ and $\delta\mathcal{E}_{RMS} = 95\ meV$.

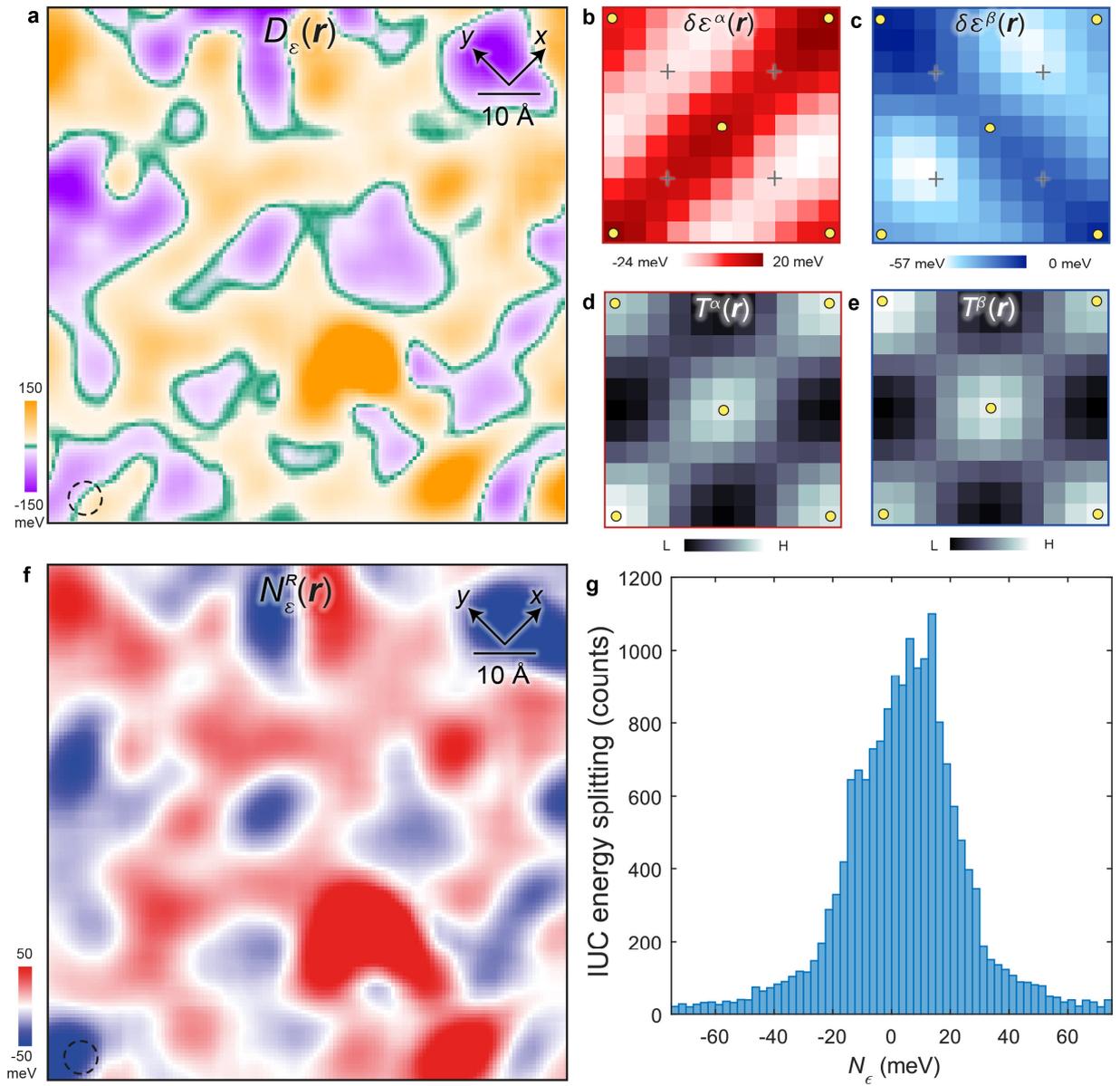

**Extended Data Fig. 8. Nematicity analysis of the same FOV in Extended Data Fig. 7.** (**a**) Analysis of the IUC charge transfer order parameter $D_\varepsilon(r)$. The area ratio $A_{red}/A_{blue}$ is 1.8. The largest domain size is ~29 nm². (**b** & **c**) unit cell averaged structure of $D_\varepsilon(r)$. (**d** & **e**) unit cell averaged structure of $T(r)$. (**f**) Image of oxygen-site-specific nematic order parameter $N_\varepsilon(r)$. (**g**) Histogram of $N_\varepsilon(r)$ whose RMS is ~28 meV.

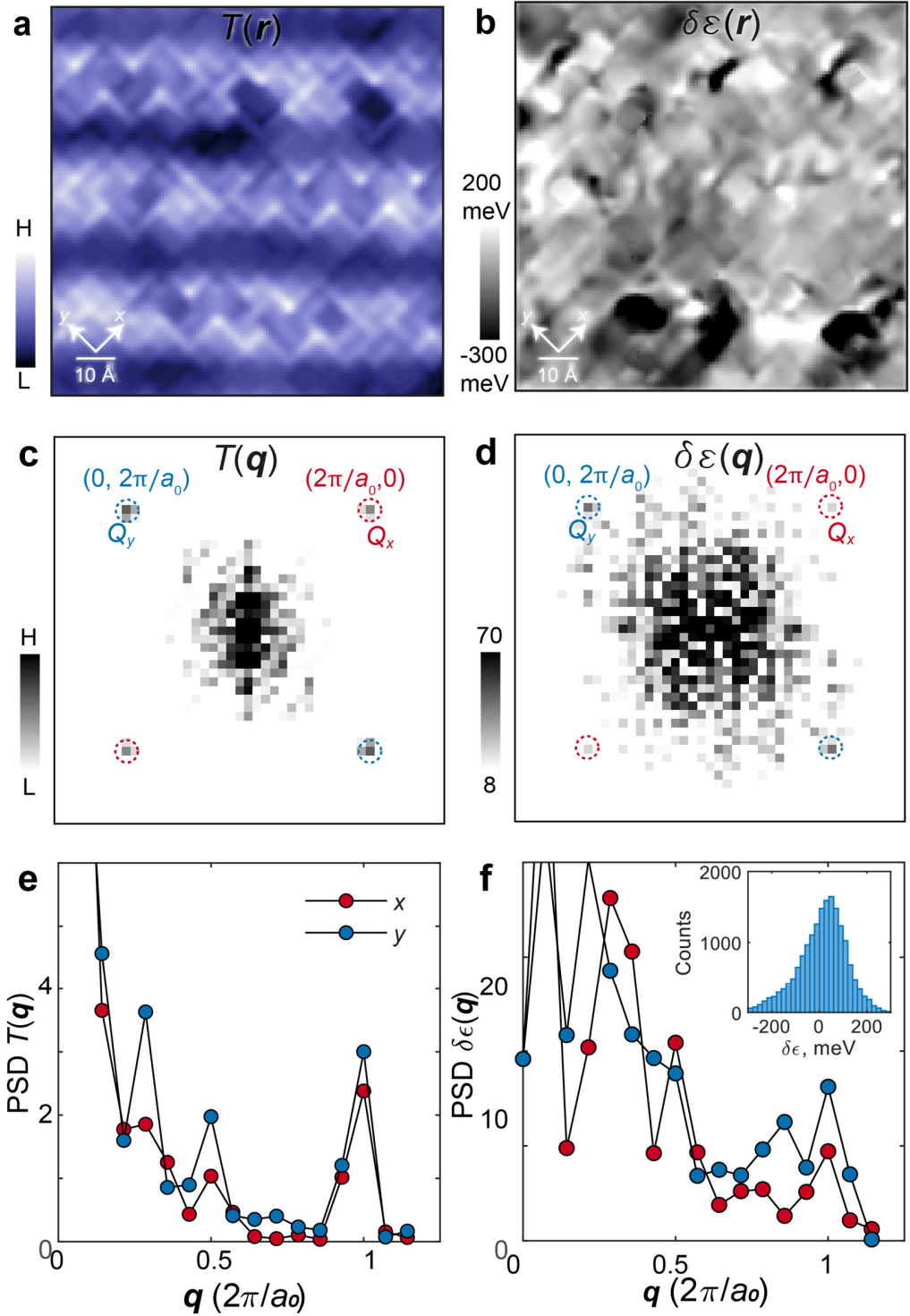

**Extended Data Fig. 9. Fourier analysis of charge transfer energy variations map $\delta\mathcal{E}(r)$ in another FOV.** (e) $T(q)$ shows the ratio of the transverse averaged intensity at the Bragg peaks is $\frac{Q_y}{Q_x} = 1.3$. (f) $\delta\mathcal{E}(q)$ shows the intensity ratio of $\frac{Q_y}{Q_x} = 1.7$ and $\delta\mathcal{E}_{RMS}$ is $\sim 129\ meV$.

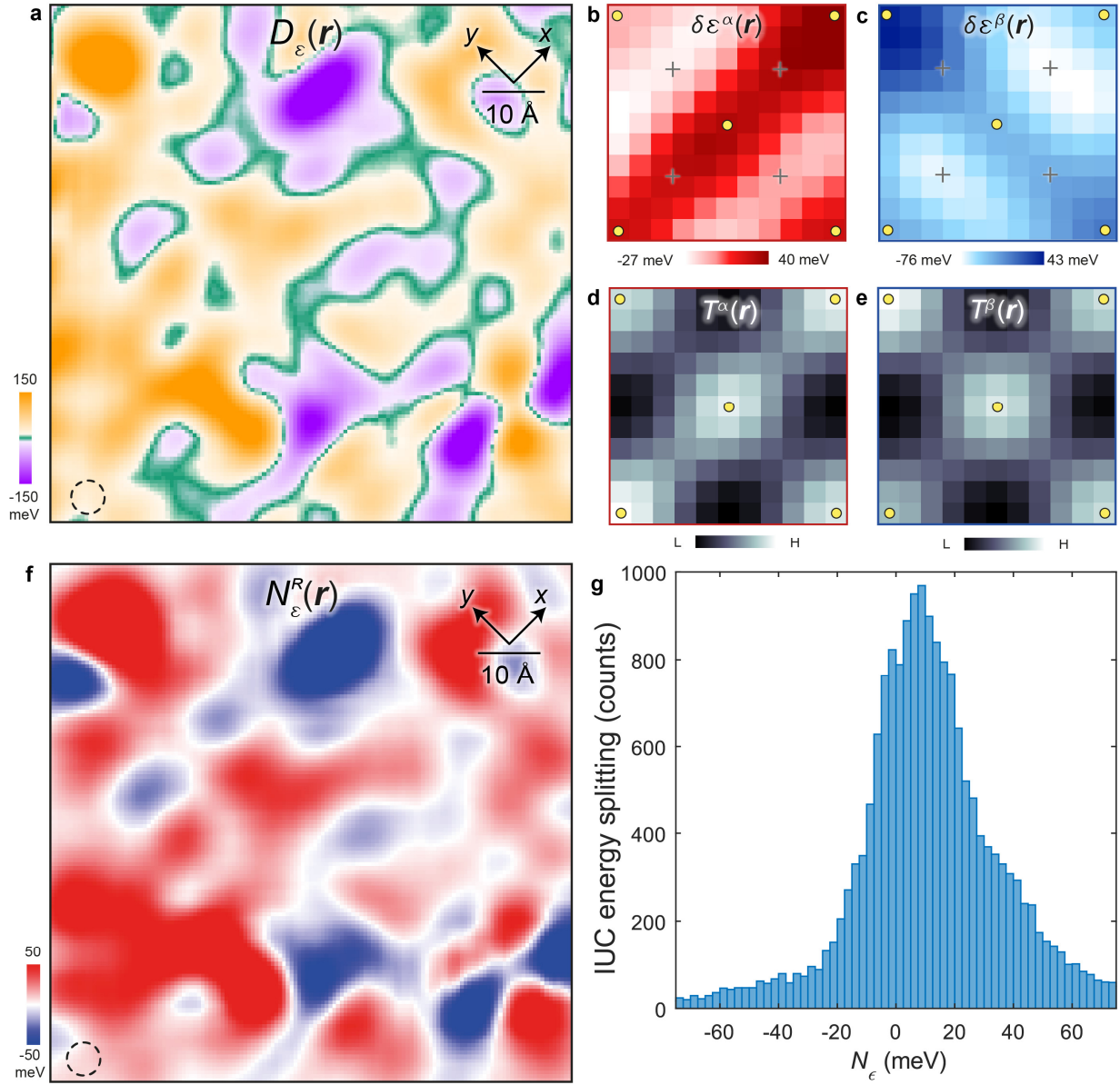

**Extended Data Fig. 10. Nematicity analysis of the same FOV in Extended Data Fig. 9.** (a) Analysis of the IUC charge transfer order parameter $D_\varepsilon(r)$. The area ratio $A_{red}/A_{blue}$ is ~2.3. The largest domain size is ~33 nm². (**b** & **c**) unit cell averaged structure of $D_\varepsilon(r)$ and (**d** & **e**) of $T(r)$. (**f**) Image of oxygen-site-specific nematic order parameter $N_\varepsilon(r)$. (**g**) Histogram of $N_\varepsilon(r)$ whose RMS is ~34 meV.

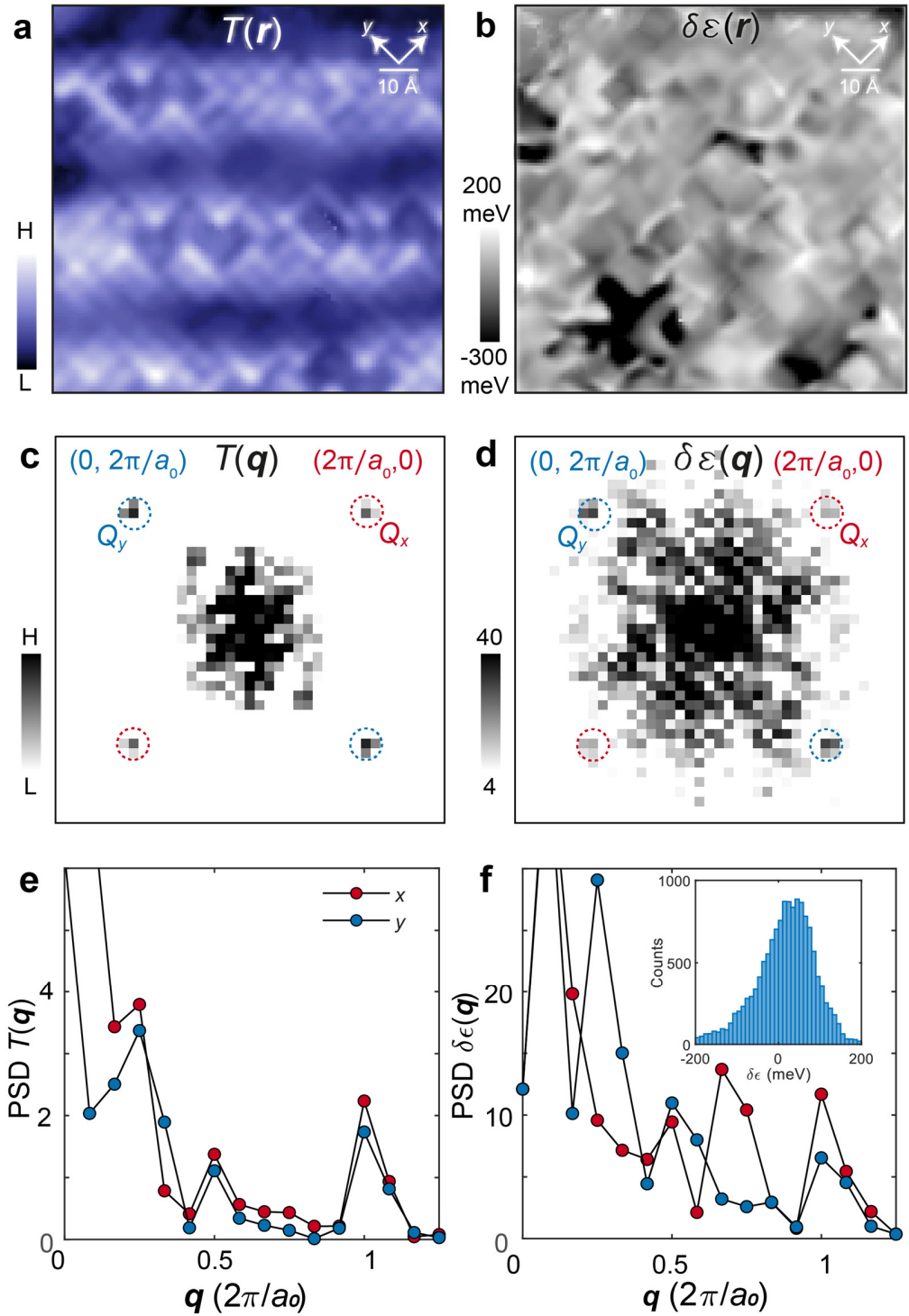

**Extended Data Fig. 11. Fourier analysis of charge-transfer energy variations map $\delta\mathcal{E}(r)$ in another FOV.** (**e**) $T(q)$ shows the ratio of the transverse averaged intensity at the Bragg peaks is $\frac{Q_y}{Q_x} = 1.3$. (**f**) $\delta\mathcal{E}(q)$ shows $\frac{Q_y}{Q_x} = 1.8$ and $\delta\mathcal{E}_{RMS} = \sim 105\ meV$.

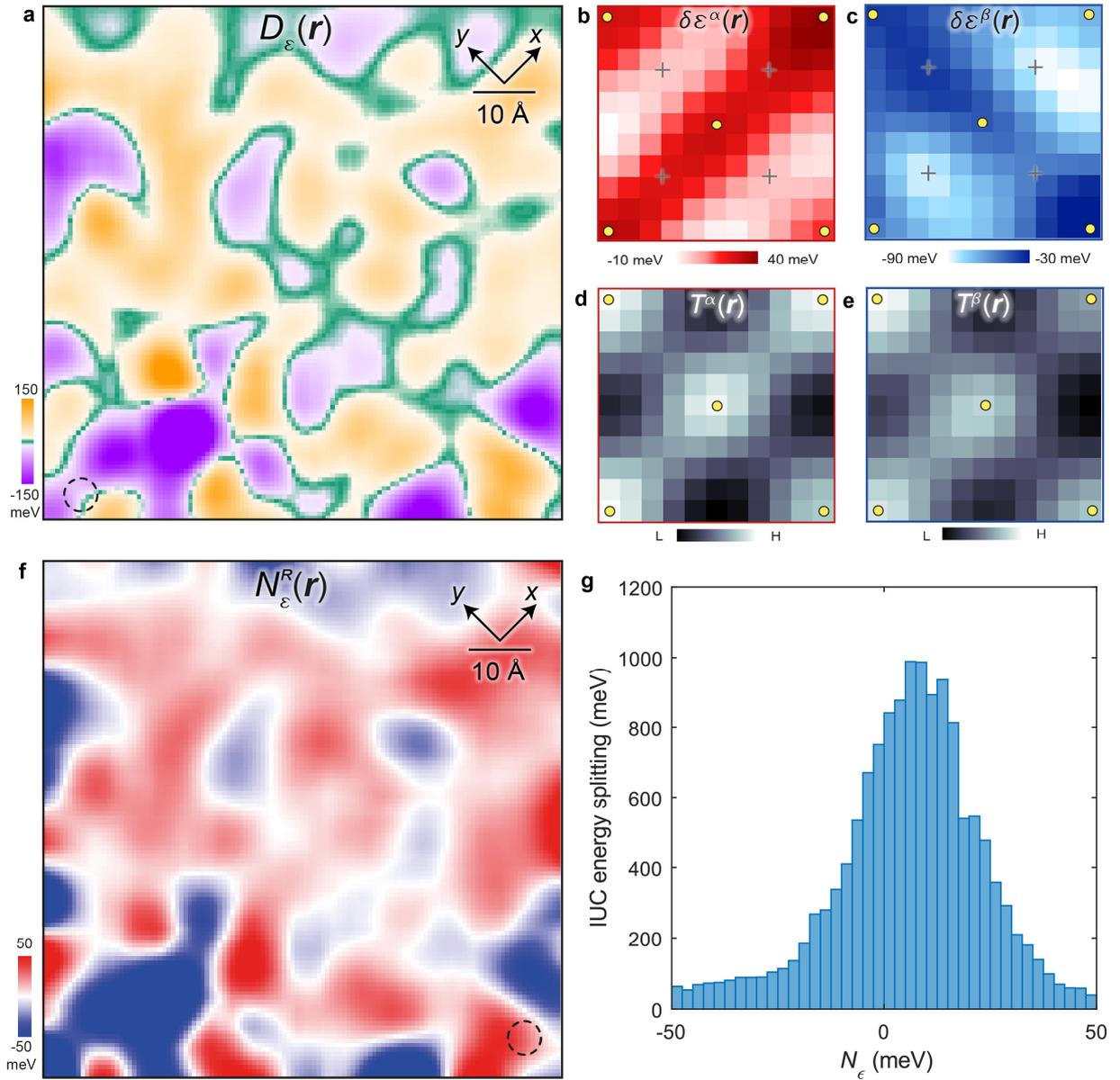

**Extended Data Fig. 12. Nematicity analysis of the same FOV in Extended Data Fig. 11.** (**a**) Analysis of the IUC symmetry charge transfer order parameter $D_\varepsilon(\mathbf{r})$. (**b** & **c**) unit cell averaged structure of $D_\varepsilon(\mathbf{r})$ and (**d** & **e**) of $T(\mathbf{r})$. (**f**) Image of oxygen-site-specific nematic order parameter $N_\varepsilon(\mathbf{r})$. The area ratio $A_{red}/A_{blue}$ is ~2.2. The largest domain size is ~23 nm². (**g**) Histogram of $N_\varepsilon(\mathbf{r})$ whose RMS is ~26 meV.

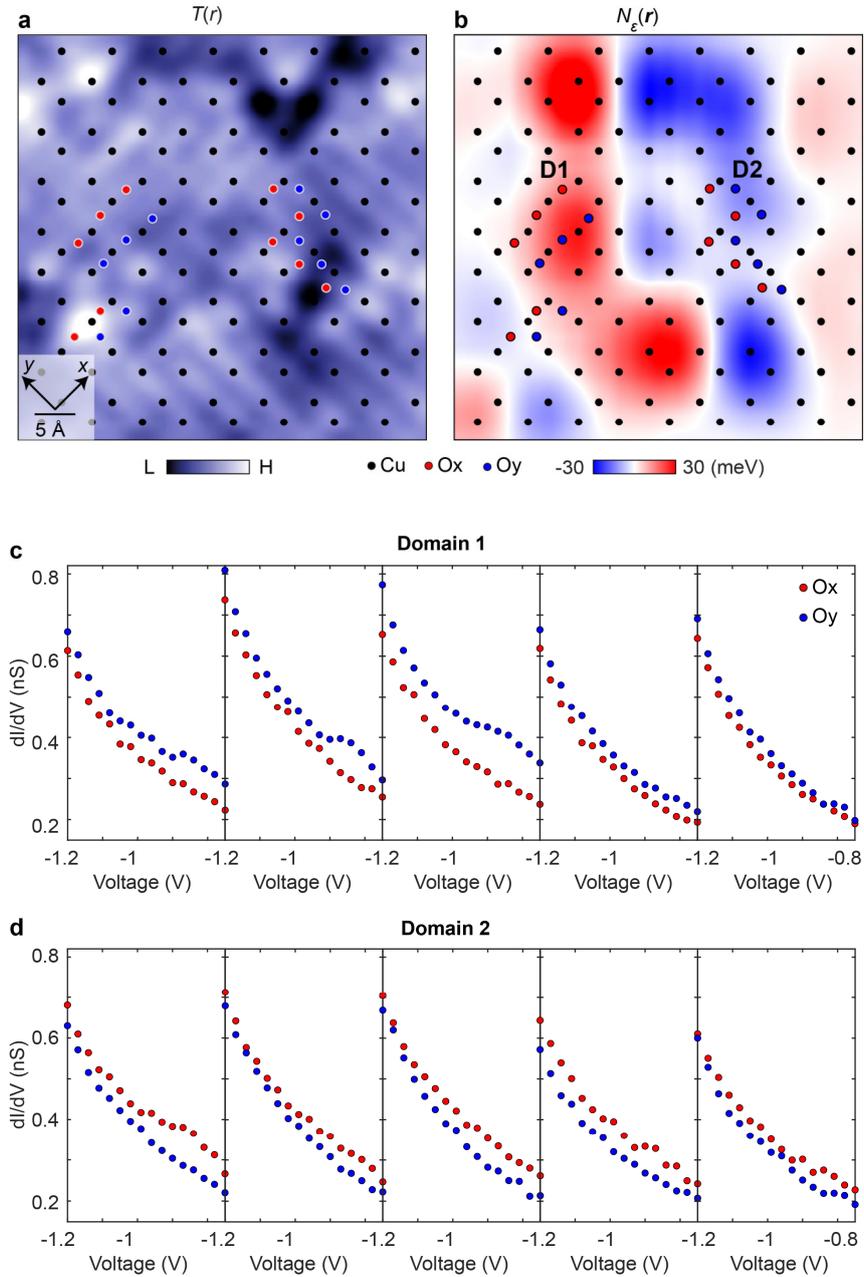

**Extended Data Fig. 13. Intra-unit-cell splitting of unprocessed differential conductance spectra (FOV1).** **(a)** Topograph of the FOV where the spectra are measured. **(b)** Image of nematic intra-unit-cell order parameter $N_\varepsilon(r)$ sampled on oxygen sites. Two Ising domains are labelled. **(c)** 10 d$I$/d$V$ point spectra sampled on the oxygen sites from Domain 1 in (b). Clearly these point spectra show the Ox spectra are shifted by -50 meV~-30 meV with respect to their intra-unit-cell Oy spectra. The locations of the Ox and Oy sites are shown as circles in (a-b). **(d)** 10 point spectra sampled on the oxygen sites from Domain 2 reveal that the Oy spectra are shifted by -50 meV~-30 meV with respect to the Ox spectra from the same unit cell. The locations of the unit cell and oxygen sites are shown in circles in (a-b). The dI/dV splitting to opposite directions in the two different domains, proves charge transfer energy splitting occurs in unprocessed differential conductance spectra.

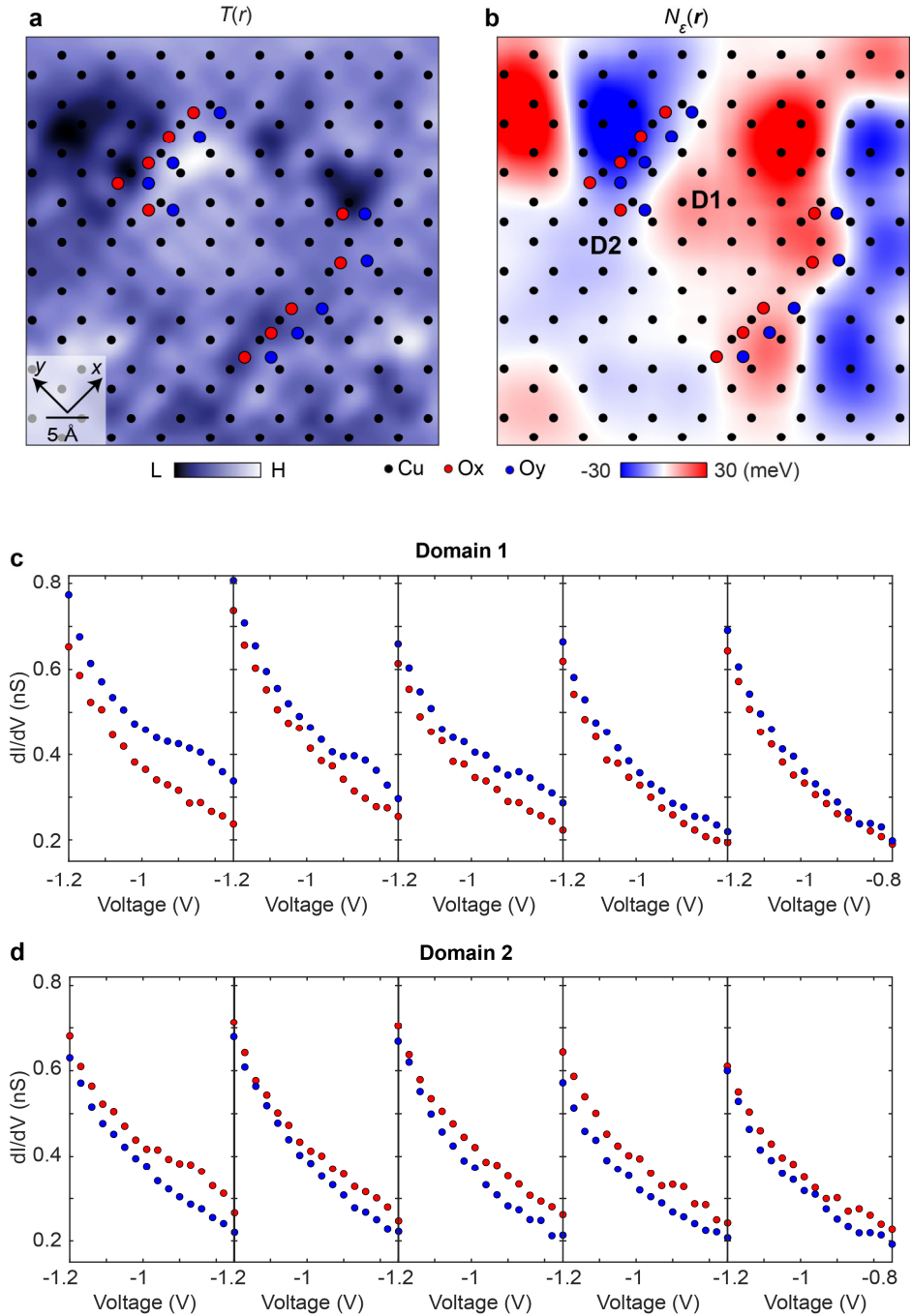

**Extended Data Fig. 14. Intra-unit-cell splitting of unprocessed differential conductance spectra (FOV2).** (a) Topograph of the FOV where the spectra are measured. (b) Image of orbital-order intra-unit-cell order parameter $N_\varepsilon(r)$ sampled on oxygen sites. Two Ising domains are labelled. (c) 10 d$I$/d$V$ point spectra sampled on the oxygen sites from Domain 1 in (b) reveal that the Ox spectra are shifted by -50 meV~-30 meV with respect to the Oy spectra from the same unit cell. (d) 10 point spectra sampled on the oxygen sites from Domain 2 reveal that the Oy spectra are shifted by -50 meV~-30 meV with respect to the Ox spectra from the same unit cell.

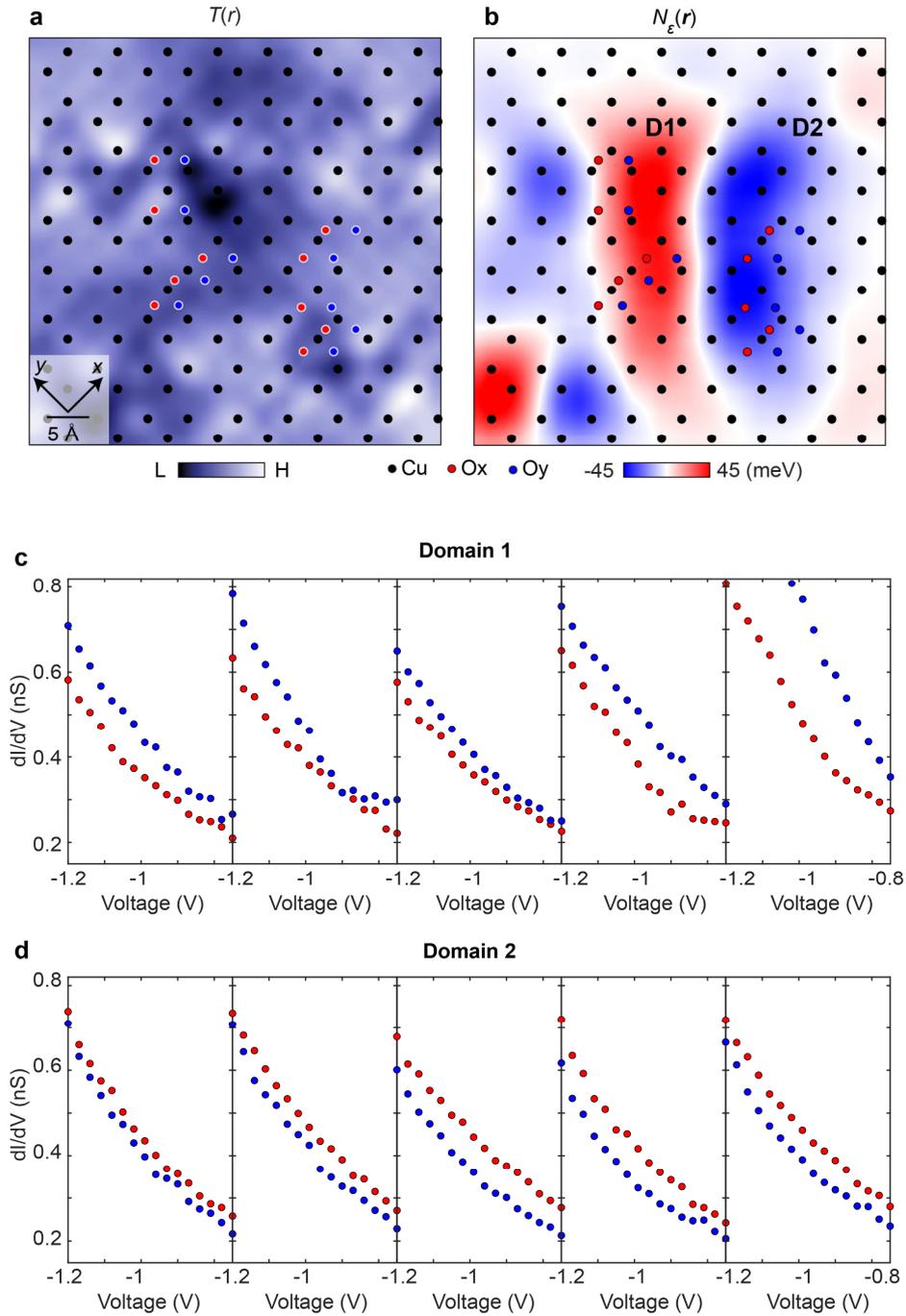

**Extended Data Fig. 15. Intra-unit-cell splitting of unprocessed differential conductance spectra (FOV3).** (a) Topograph of the FOV where the spectra are measured. (b) Image of orbital-order intra-unit-cell order parameter $N_\varepsilon(\mathbf{r})$ sampled on oxygen sites. Two Ising domains are labelled. (c) 10 d$I$/d$V$ point spectra sampled on the oxygen sites from Domain 1 in reveal that the Ox spectra are shifted by -50 meV~-30 meV with respect to the Oy spectra from the same unit cell in (b). (d) 10 point spectra sampled on the oxygen sites from Domain 2 reveal that the Oy spectra are shifted by -50 meV ~-30 meV with respect to the Ox spectra from the same unit cell.

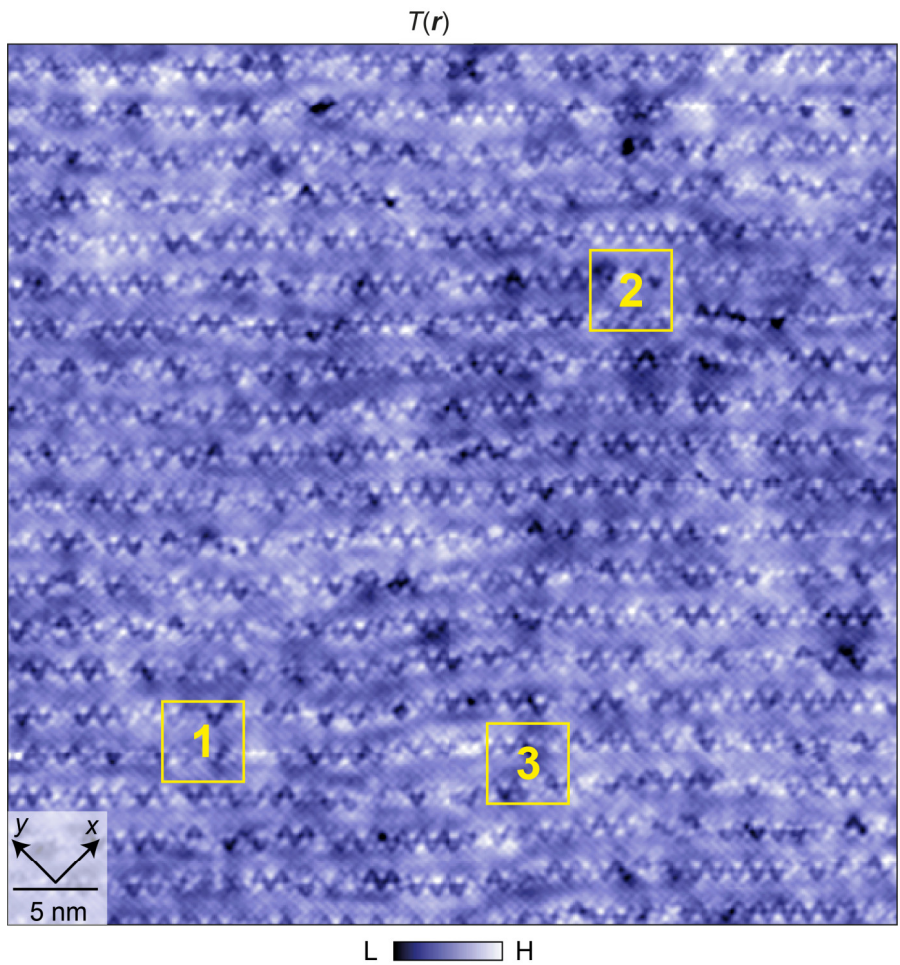

**Extended Data Fig. 16. Large FOV topography showing the randomly chosen regions of above experiments.** FOVs 1-3 in this topography corresponds to the FOVs in Figs. 13-15, respectively. $T = $ 4.2 K, $V_S$ = -750 mV and $I_S$ = 25 pA.

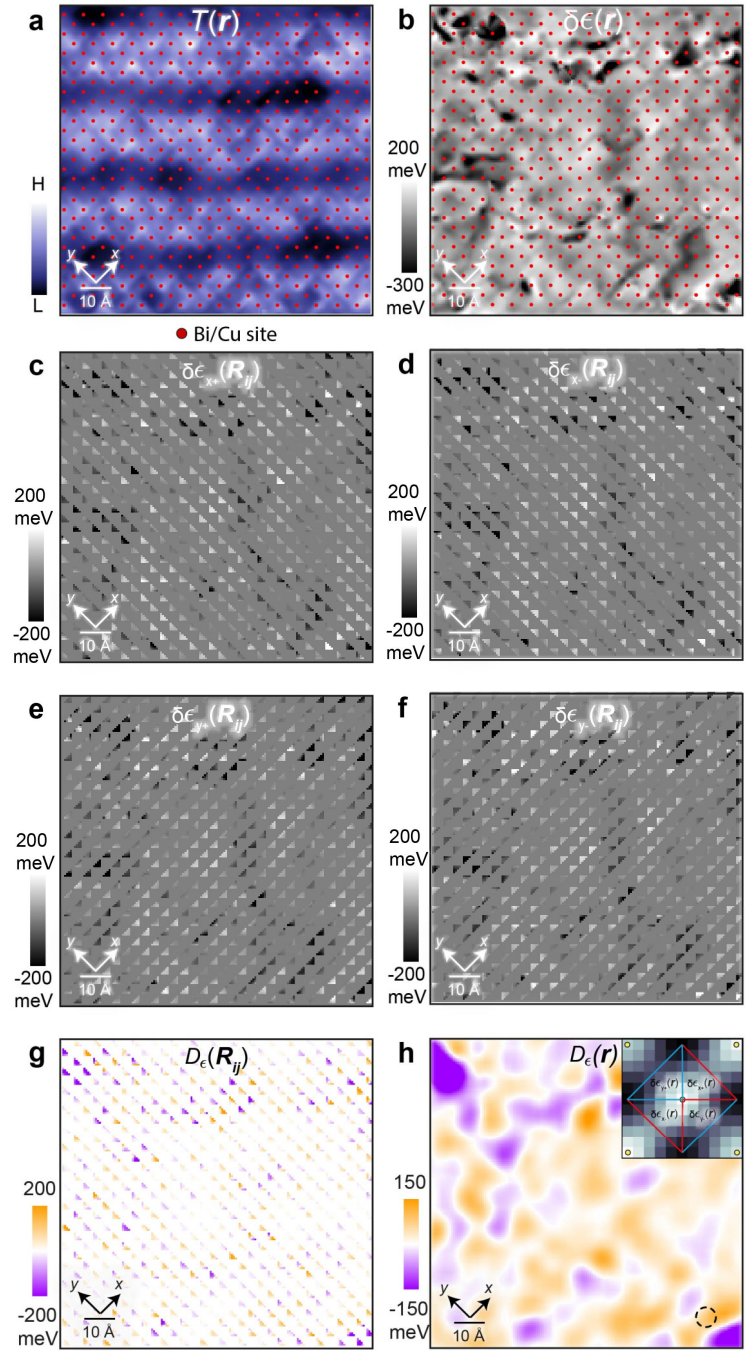

**Extended Data Fig. 17. Visualizing Intra-unit-cell Charge Transfer Order Parameter.** (**a**) identification of the location of the Bi/Cu sites as red circles in $T(r)$. (**b**) Apply the Cu coordinates to $\delta\mathcal{E}(r)$. (**c** – **f**) decompose each CuO$_2$ unit cell in $\delta\mathcal{E}(r)$ into four quadrants including $\delta\varepsilon_{x+}$, $\delta\varepsilon_{x-}$, $\delta\varepsilon_{y+}$ and $\delta\varepsilon_{y-}$. (**g**) The d-symmetry IUC order parameter $D_\varepsilon(R_{i,j})$. (**h**) $D_\varepsilon(r)$ image calculated by applying a gaussian smoothing of filter size $r$ = 3.2 Å to $D_\varepsilon(R_{i,j})$, with the inset describing the quadrant masks used.

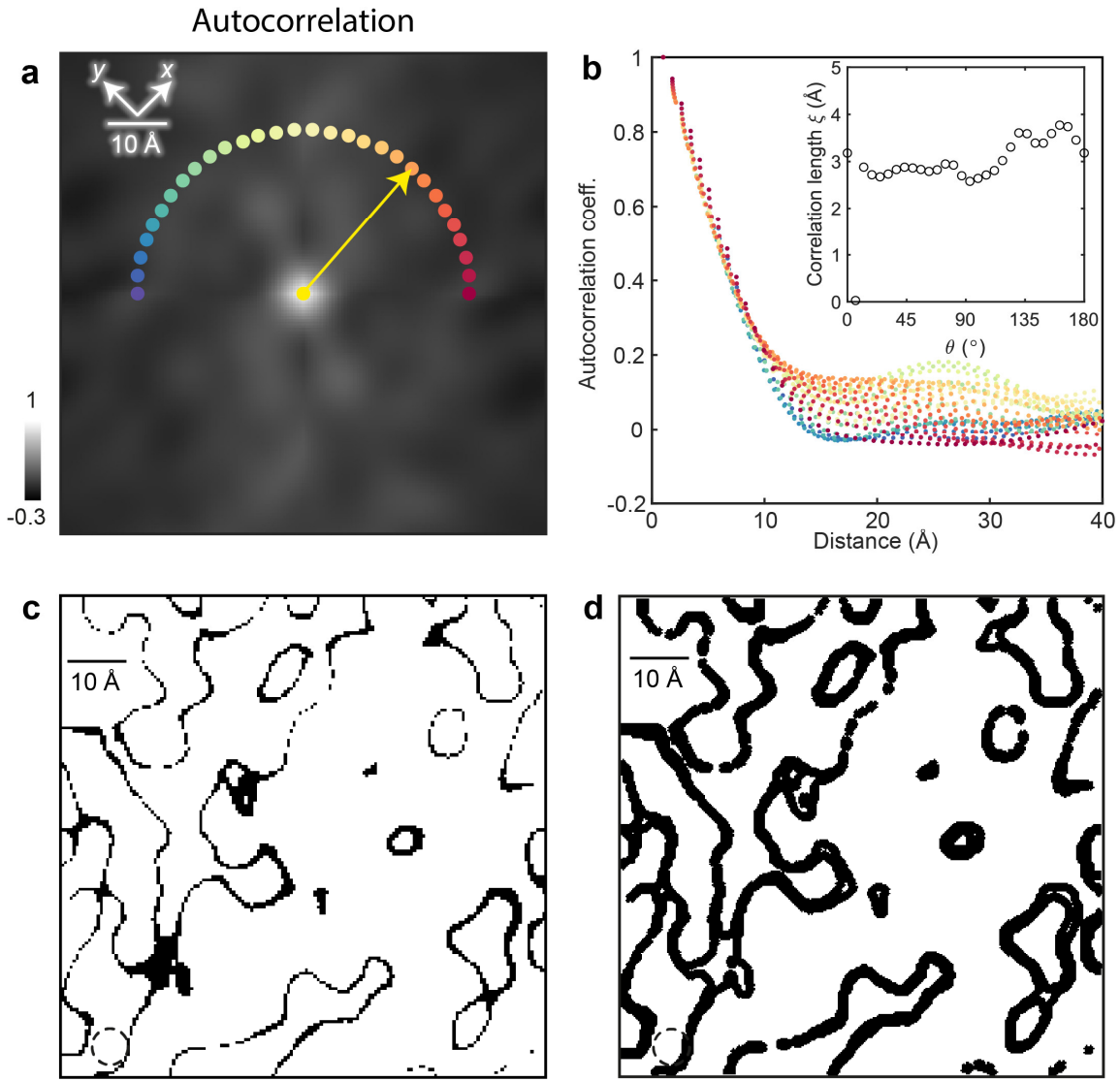

**Extended Data Fig. 18. Area Fraction of Charge Quadrupoles.** (**a**) Auto-correlation of $N_\varepsilon(r)$ map shown in Fig. 3a of the main text. (**b**) Radial plot of the autocorrelation function where the colours relate to the dots at different angles shown in (a). Each linecut is taken from the center of the AC image to the dots at various angles. The inset here is correlation length $\xi$ as a function of radial angle, the radial average of this correlation length $\bar{\bar{\xi}} = 3.1$ Å. (**c**) Location of the domain boundaries determined from Fig. 3a of the main text. (**d**) The domain walls broadened by $\bar{\bar{\xi}}$.

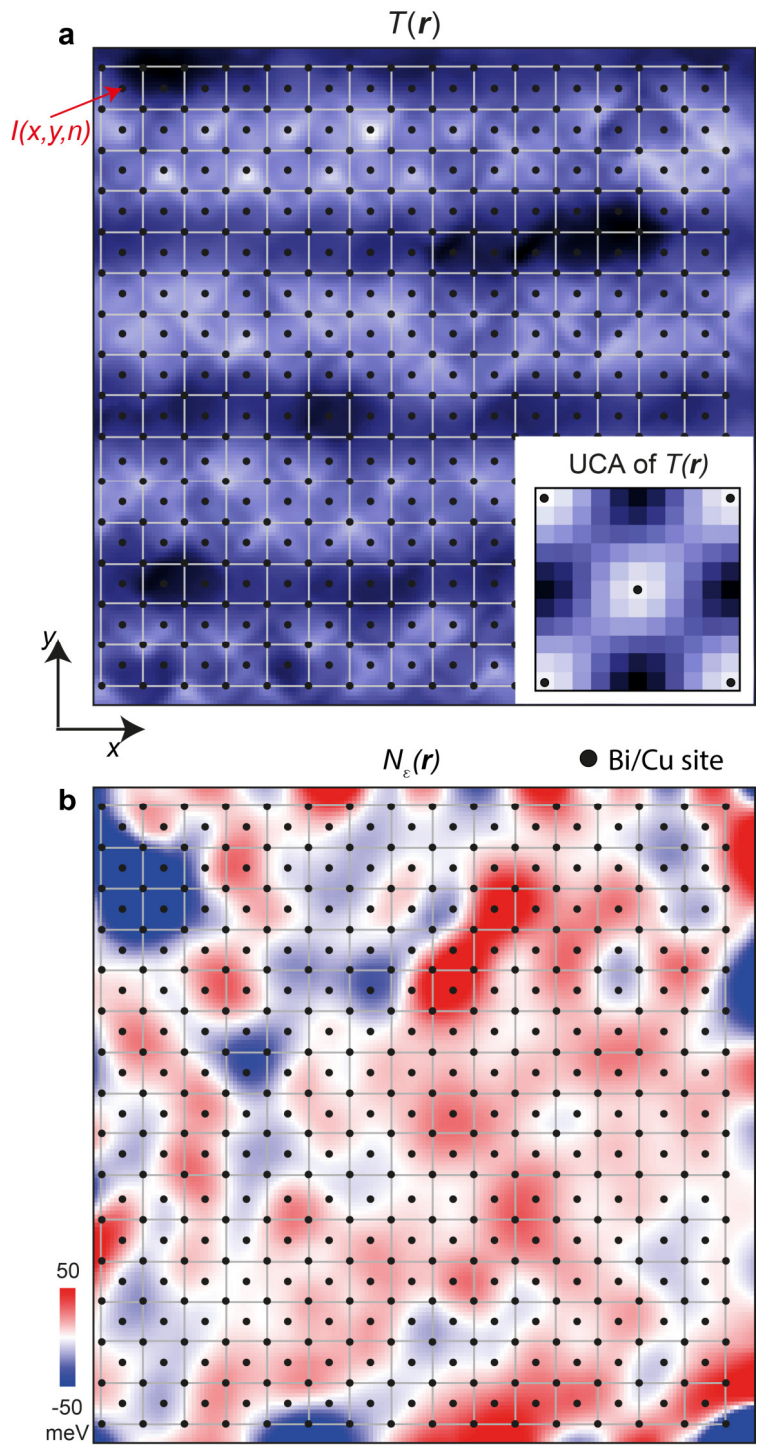

**Extended Data Fig. 19. Unit-cell averaging (UCA) analysis.** (**a**) Identification of the Cu site in the topograph $T(r)$. Each unit cell is defined by a Cu site (black circles) in the center and four Cu sites at the corners. The averaged image of all the unit cells is a UCA image presented in the inset. (**b**) The CuO$_2$ unit cells are categorized into two zones of Ising domains from the parameter $N_\varepsilon(r)$ map. The UCA images of each Ising domain is subsequently calculated.

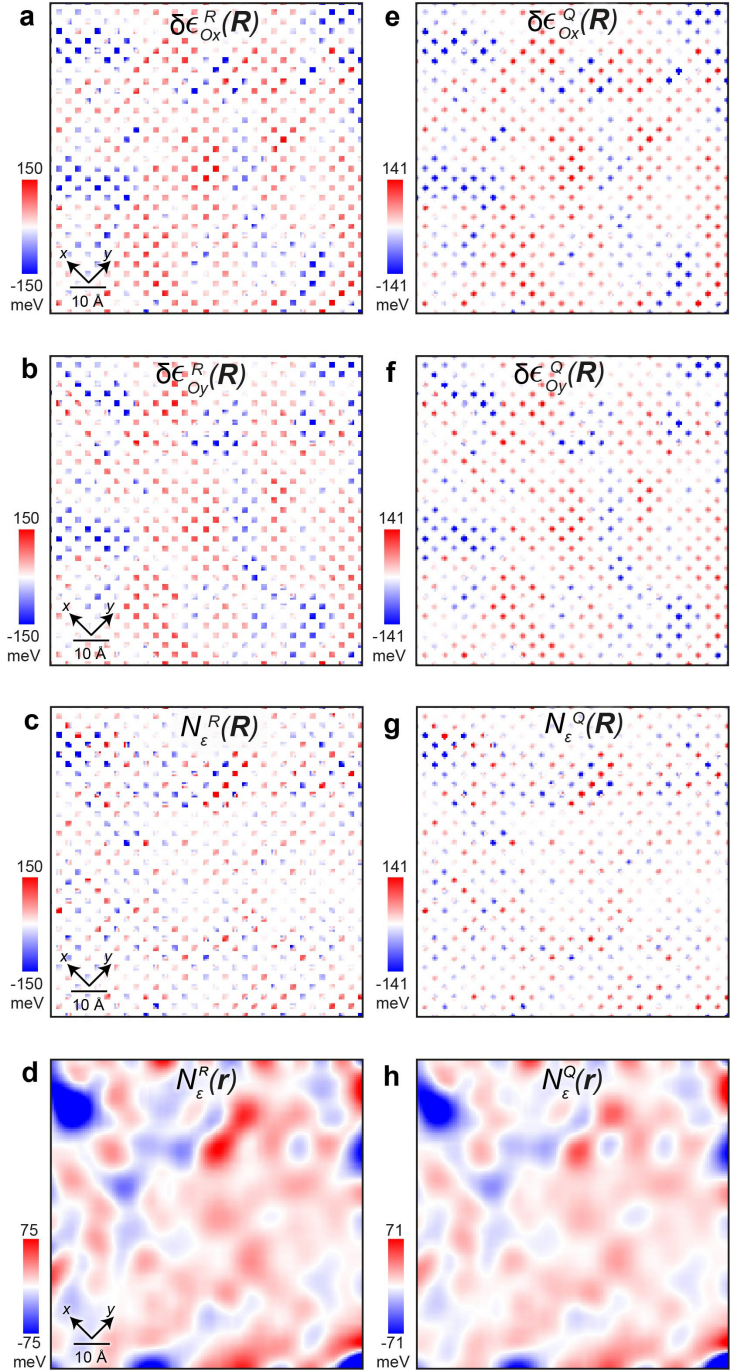

**Extended Data Fig. 20. Oxygen site-specific decomposition of** $\delta\mathcal{E}(r)$. The left column shows the measurements from the real space (R) and the right column displays the measurements from the reciprocal space (Q). (**a** & **e**) The value of $\delta\mathcal{E}$ is measured at every $O_x$ site and show the resulting function $\delta\mathcal{E}^R_{Ox}(R_{i,j})$ and $\delta\mathcal{E}^Q_{Ox}(R_{i,j})$. (**b** & **f**) $\delta\mathcal{E}$ measured at each $O_y$ site, resulting in the images of $\delta\mathcal{E}^R_{Oy}(R_{i,j})$ and $\delta\mathcal{E}^Q_{Oy}(R_{i,j})$. (**c** & **g**) Oxygen-site-specific nematic order parameter $N_\mathcal{E}(R_{i,j})$ is defined as the difference between $\delta\mathcal{E}_{Ox}(R_{i,j})$ and $\delta\mathcal{E}_{Oy}(R_{i,j})$. (**d** & **h**) Spatially averaged nematic order parameter $N_\mathcal{E}(r)$. The measurements from the *r*-space and the *q*-space are almost identical.

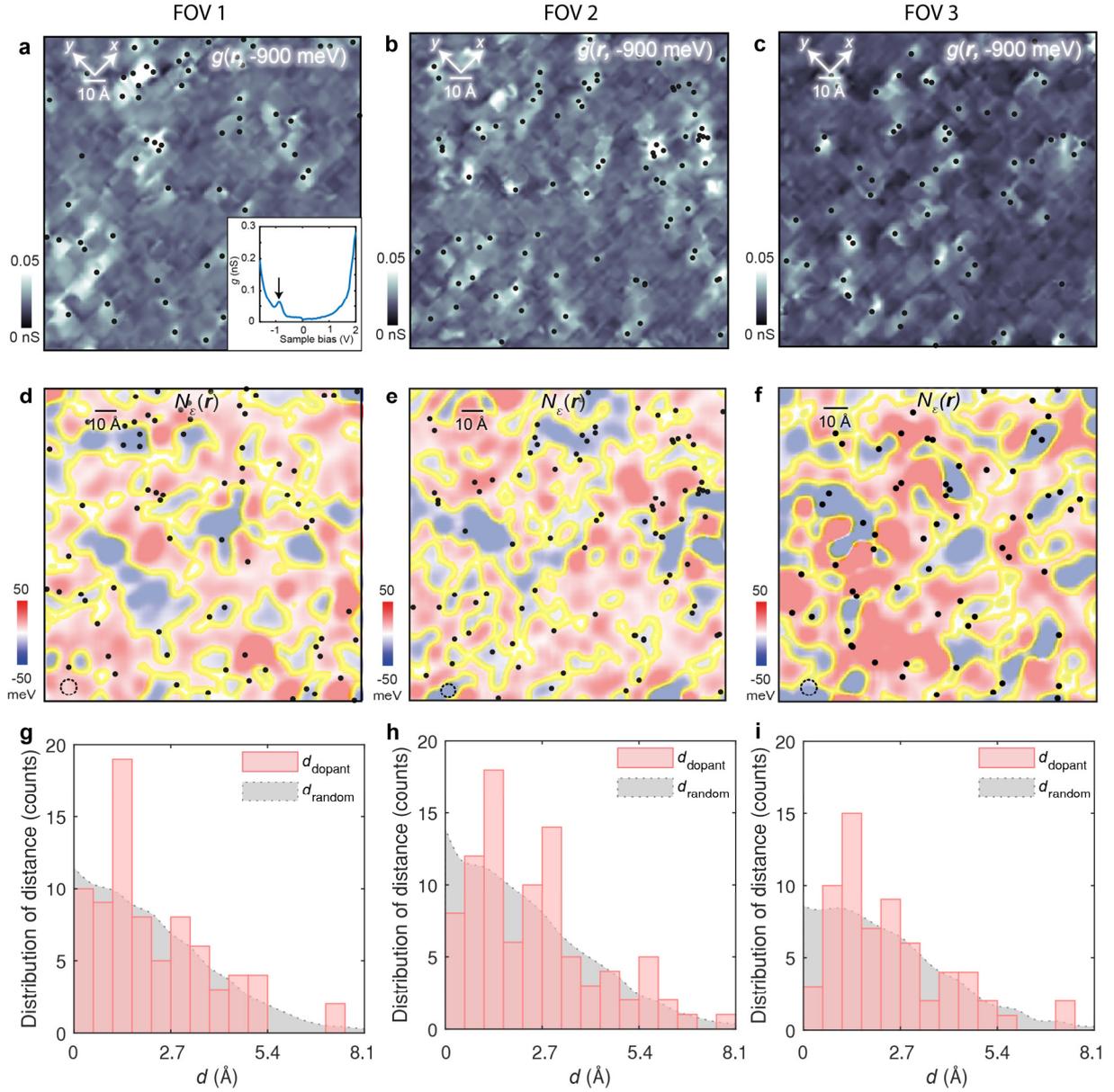

**Extended Data Fig. 21. Statistical validation of dopant oxygen ion pinning of orbital ordered domains.** (**a-c**) Three independent differential conductance maps $g(r, -900\,\text{meV})$ where the locations of the oxygen dopants are identified as black circles. Inset of (a) shows a typical dI/dV spectrum of an oxygen dopant. (**d-f**) Oxygen-specific order parameter $N_\varepsilon(r)$ with overlaying oxygen dopants as black circles. The $N_\varepsilon(r)$ are measured simultaneously as $g(r, -900\,\text{meV})$ in (**a-c**), where the domain walls are highlighted. (**g-i**) The $d_{\text{dopant}}$ histogram (pink bar) is the distance from each dopant to the nearest location on the domain walls. The $d_{\text{random}}$ histogram (grey curve) is the expectations of the distance between simulated random points and its nearest point in the domain walls.

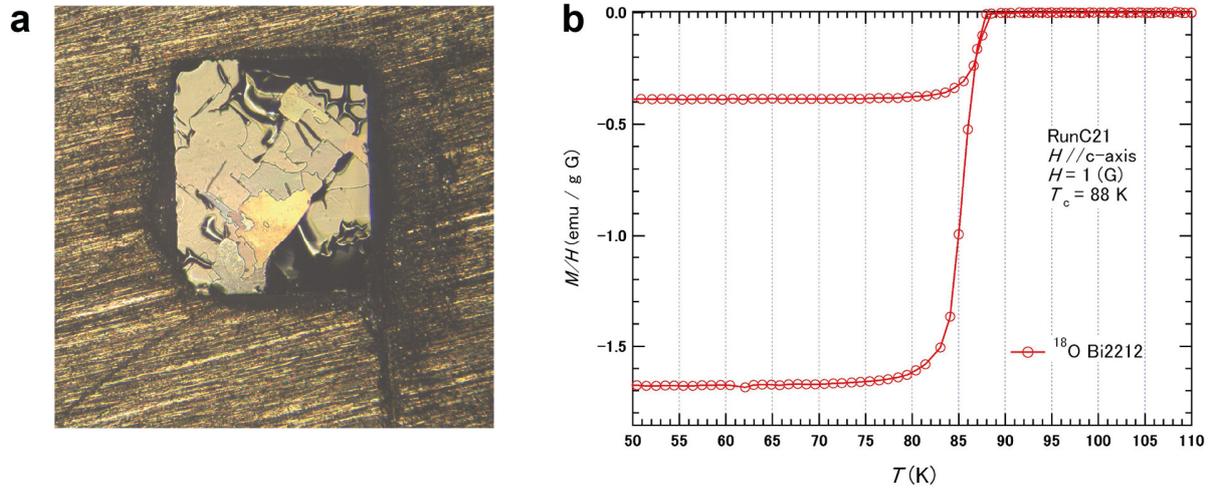

**Extended Data Fig. 22. Characterization of $Bi_2Sr_2CaCu_2O_{8+x}$ sample.** (**a**) A photo of a single crystal of $Bi_2Sr_2CaCu_2O_{8+x}$ studied throughout the paper. (**b**) Plot of the magnetic susceptibility of the sample from $T$ = 50 K to 110 K, with the superconducting transition clearly seen at $T_c = 88$ K.

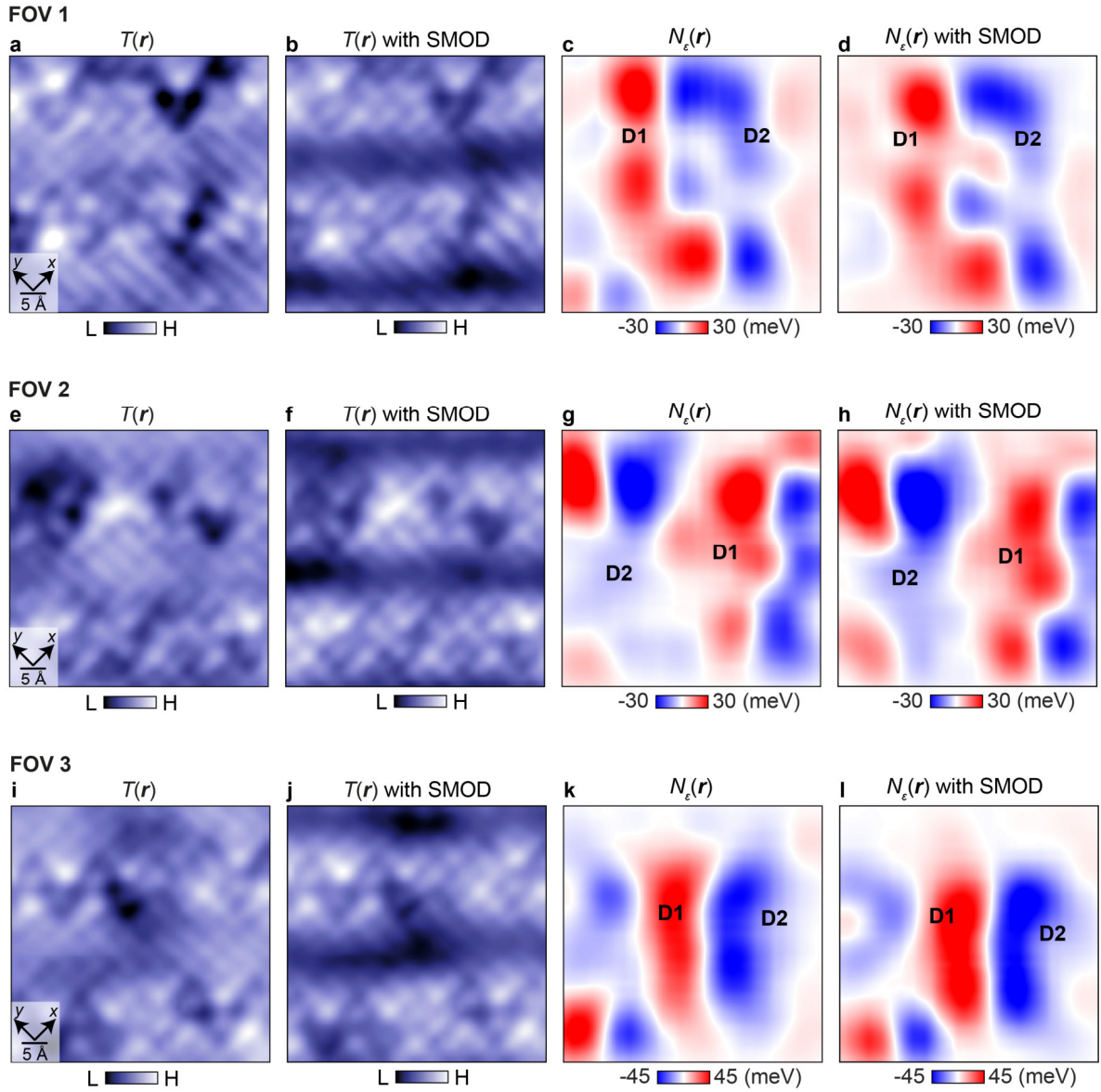

**Extended Data Fig. 23. Supermodulation does not affect orbital ordering domains. (a, e, i)** show the topographs where the supermodulation is filtered. **(b, f, j)** present the topographs where the supermodulation is kept. **(c, g, k)** show the orbital ordering domains calculated from the topographs in the first column where the supermodulation is removed. **(d, h, l)** show the orbital ordering domains calculated from the topographs in the second columns where the supermodulation remains. The orbital ordering domains from the third (without supermodulation) and fourth columns (with supermodulation) are virtually identical. The supermodulation has virtually no effect on the orbital-order parameter used in this paper.